\documentclass[twocolumn]{aastex631}

\newcommand{\planet}{HD 206893 B}

\newcommand{\plusminus}[2]{$^{+#1}_{-#2}$}
\newcommand{\breads}{\texttt{breads}}

\usepackage{color}
\usepackage{amsmath}
\usepackage{amssymb}
\usepackage{xcolor}
\usepackage{mathrsfs}
\usepackage{natbib}
\usepackage{gensymb}
\usepackage{xspace}
\usepackage{subfigure}
\usepackage{placeins}
\usepackage{array}
\usepackage{CJK}

\newcommand{\Rj}{\ensuremath{R_{\rm{Jup}}}\xspace}
\newcommand{\Mj}{\ensuremath{M_{\rm{Jup}}}\xspace}

\newcommand{\Teff}{\ensuremath{T_\mathrm{eff}}\xspace}
\newcommand{\logg}{\ensuremath{\log{g}}\xspace}

\newcommand{\vsini}{\ensuremath{v \sin i}\xspace} 
\newcommand{\angstrom}{\textup{\AA}}

\newcommand{\spin}{$v \sin{i}$}
\newcommand{\kband}{\textit{K-}band\xspace}
\newcommand{\jband}{\textit{J-}band\xspace}
\newcommand{\hband}{\textit{H-}band\xspace}
\newcommand{\prt}{\texttt{petitRADTRANS}\xspace}
\newcommand{\caltech}{Department of Astronomy, California Institute of Technology, Pasadena, CA 91125, USA}
\newcommand{\gps}{Division of Geological \& Planetary Sciences, California Institute of Technology, Pasadena, CA 91125, USA}
\newcommand{\ucsc}{Department of Astronomy \& Astrophysics, University of California, Santa Cruz, CA95064, USA}
\newcommand{\keck}{W. M. Keck Observatory, 65-1120 Mamalahoa Hwy, Kamuela, HI, USA}
\newcommand{\ucla}{Department of Physics \& Astronomy, 430 Portola Plaza, University of California, Los Angeles, CA 90095, USA}
\newcommand{\jpl}{Jet Propulsion Laboratory, California Institute of Technology, 4800 Oak Grove Dr.,Pasadena, CA 91109, USA}
\newcommand{\ucsd}{Department of Astronomy and Astrophysics, University of California, San Diego, La Jolla, CA 92093}
\newcommand{\ucsdphys}{Department of Physics, University of California, San Diego, La Jolla, CA 92093}

\newcommand{\osu}{Department of Astronomy, The Ohio State University, 100 W 18th Ave, Columbus, OH 43210 USA}

\newcommand{\arizona}{James C. Wyant College of Optical Sciences, University of Arizona, Meinel Building 1630 E. University Blvd., Tucson, AZ. 85721}

\newcommand{\carnegiew}{Earth and Planets Laboratory, Carnegie Institution for Science, Washington, DC, 20015}

\newcommand{\ua}{Lunar and Planetary Laboratory, University of Arizona, Tucson, AZ 85721, USA}

\usepackage{CJK}
\usepackage{bm} 
\usepackage{amsmath}
\usepackage{multirow}
\usepackage{gensymb}

\shorttitle{HD 206893 B at High Spectral Resolution using KPIC/NIRSPEC}
\shortauthors{Sappey et al.}

\graphicspath{{./}{}}

\begin{document}
\begin{CJK*}{UTF8}{gbsn}
\title{HD 206893 B at High Spectral Resolution with the Keck Planet Imager and Characterizer (KPIC)}

\author[0000-0003-1399-3593]{Ben Sappey}
\affiliation{\ucsdphys}

\author[0000-0002-9936-6285]{Quinn Konopacky}
\affiliation{\ucsd}

\author[0000-0001-5173-2947]{Clarissa R. Do \'O}
\affiliation{\ucsdphys}

\author{Travis Barman}
\affiliation{\ua}

\author[0000-0003-2233-4821]{Jean-Baptiste Ruffio}
\affiliation{\ucsd}

\author[0000-0003-0774-6502]{Jason Wang (王劲飞)}
\affiliation{Center for Interdisciplinary Exploration and Research in Astrophysics (CIERA) and Department of Physics and Astronomy, Northwestern University, Evanston, IL 60208, USA}

\author[0000-0002-9807-5435]{Christopher A. Theissen}
\affiliation{\ucsd}

\author[0000-0002-1392-0768]{Luke Finnerty}
\affiliation{\ucla}

\author{Jerry Xuan}
\affiliation{\caltech}

\author{Katelyn Hortsman}
\affiliation{\caltech}
\altaffiliation{NSF Graduate Research Fellow}

\author{Dimitri Mawet}
\affiliation{\caltech}
\affiliation{\jpl}

\author[0000-0003-0097-4414]{Yapeng Zhang}
\affiliation{\caltech}

\author[0000-0001-9164-7966]{Julie Inglis}
\affiliation{\gps}

\author[0000-0003-0354-0187]{Nicole L. Wallack}
\affiliation{\carnegiew}

\author[0000-0002-1838-4757]{Aniket Sanghi}
\affiliation{\caltech}


\author{Ashley Baker}
\affiliation{\caltech}

\author{Randall Bartos}
\affiliation{\jpl}

\author[0000-0003-0787-1610]{Geoffrey A. Blake}
\affiliation{Division of Geological and Planetary Sciences, California Institute of Technology, MC 150-21, Pasadena, CA 91125, USA}

\author{Charlotte Z. Bond}
\affiliation{UK Astronomy Technology Centre, Royal Observatory, Edinburgh EH9 3HJ, United Kingdom}

\author[0000-0003-4737-5486]{Benjamin Calvin}
\affiliation{\caltech}
\affiliation{\ucla}

\author{Sylvain Cetre}
\affiliation{\keck}

\author[0000-0001-8953-1008]{Jacques-Robert Delorme}
\affiliation{\keck}

\author{Greg Doppmann}
\affiliation{\keck}

\author{Daniel Echeverri}
\affiliation{\caltech}

\author[0000-0002-0176-8973]{Michael P. Fitzgerald}
\affiliation{\ucla}

\author[0000-0002-5370-7494]{Chih-Chun Hsu}
\affil{Center for Interdisciplinary Exploration and Research in Astrophysics (CIERA), Northwestern University,
1800 Sherman Ave, Evanston, IL 60201, USA}

\author[0000-0001-5213-6207]{Nemanja Jovanovic}
\affiliation{\caltech}

\author[0000-0002-4934-3042]{Joshua Liberman}
\affiliation{\caltech}
\affiliation{\arizona}

\author[0000-0002-2019-4995]{Ronald A. L\'opez}
\affiliation{\ucla}

\author[0000-0002-0618-5128]{Emily C. Martin}
\affiliation{\ucsc}

\author{Evan Morris}
\affiliation{\ucsc}

\author{Jacklyn Pezzato-Rovner}
\affiliation{\caltech}

\author[0000-0001-5610-5328]{Caprice L. Phillips}
\affiliation{\osu}
 
\author[0000-0003-4769-1665]{Garreth Ruane}
\affiliation{\jpl}

\author{Tobias Schofield}
\affiliation{\caltech}

\author{Andrew Skemer}
\affiliation{\ucsc}

\author{Taylor Venenciano}
\affiliation{Physics and Astronomy Department, Pomona College, 333 N. College Way, Claremont, CA 91711, USA}

\author[0000-0001-5299-6899]{J. Kent Wallace}
\affiliation{\jpl}

\author[0000-0002-4361-8885]{Ji Wang (王吉)}
\affiliation{Department of Astronomy, The Ohio State University, 100 W 18th Ave, Columbus, OH 43210 USA}

\author{Peter Wizinowich}
\affiliation{\keck}

\author[0000-0002-6171-9081]{Yinzi Xin}
\affiliation{\caltech}

\begin{abstract} 
We present an atmospheric characterization and orbital analysis of HD 206893 B, an exceptionally red, L/T-transition substellar companion in a multiplanetary system, via Keck Planet Imager and Characterizer (KPIC) high-resolution (R $\sim$ 35,000) \kband spectroscopy. Using \texttt{PHOENIX} atmospheric models in a forward-model framework that fits the spectrum of the companion and diffracted starlight simultaneously, we detect HD 206893 B at $>8\sigma$ significance via cross-correlation in two epochs. We find an effective temperature for the companion of 1634\plusminus{72}{38} K and a \logg of 4.55\plusminus{0.17}{0.22}. Only accounting for statistical uncertainties, we measure the carbon-oxygen ratio (C/O) of this companion to be $0.57 \pm 0.02$, or near-solar while assuming solar metallicity. The C/O ratio we measure fits the tentative trend of $>4$ \Mj companions having near-solar C/O ratios while less massive companions have greater-than-solar C/O ratios. Using substellar evolution models, we find an age of 112\plusminus{36}{22} Myr, a mass of 22.7\plusminus{2.5}{1.7} \Mj, and a radius of $1.11 \pm0.03$\,\Rj for this companion. We also use KPIC radial velocity data to fit the orbit of HD 206893 B and analyze the orbital stability of this system. We find that the orbital stability is relatively independent of the mass of HD 206893 B, and favors an orbital configuration where B and its interior planetary companion, HD 206893 c, are co-planar. The measured C/O ratio coupled with the current architecture of the system cannot rule out a core accretion scenario, nor a disk fragmentation scenario regarding the formation pathway of HD 206893 B.


\end{abstract}



\section{Introduction} \label{sec:intro}

\par The latest iteration of instrumentation coupled to adaptive optics, designed with exoplanet characterization in mind, e.g., Gemini Planet Imager \citep{Macintosh_gpi_2008}, VLT/SPHERE \citep{Beuzit_SPHERE_2008,VLT_SPHERE_Beuzit_2019}, can directly image exoplanets and substellar companions that are young, bright, and orbit their hosts at large separations ($\sim$10--1000 au). This regime of exoplanets can be characterized via infrared spectroscopy, which is critical to uncovering the properties of these planetary systems. Determining physical parameters such as temperature, mass, radial velocity, and chemical composition can inform formation history, evolutionary state, and orbital stability. Spectrographs that have a resolving power high enough (R$\geq$10,000) can resolve individual absorption lines such as the ro-vibrational bandhead of CO at $\sim 2.3 \mu$m. Increasing the resolving power to R$\geq$30,000 means a spectrograph is also capable of measuring the projected spin velocity of a substellar companion (\spin) in addition to atmospheric composition, temperature and pressure profiles, and radial velocity. 

\subsection{Motivation}
\par As of 2024 May 1, per the NASA Exoplanet Archive\footnote{https://exoplanetarchive.ipac.caltech.edu/index.html}, 70 companions have been discovered via direct imaging, yet only 5 systems contain more than one giant planet/brown dwarf companion that can be characterized via direct imaging. Systems with multiple companions, especially at large separations and high masses, provide important data points for our understanding of both planet and system formation. Systems like HR 8799, discovered to host multiple large Jovian planets \citep{HR8799_discovery_marois_2009}, have been extensively characterized via direct imaging (e.g., \citealt{Marois_hr8799e_imaging,galicher_mband_hr8799}), integral field spectroscopy (e.g., \citealt{Konopacky_2013_CO_H2O, Ruffio_2019_OSIRIS_MRS_HR8799,bowler_hr8799_2010_osiris,Greenbaum_hr8799_gpi_2018}), and slit spectroscopy at both moderate and high resolution (e.g., \citealt{Wang_HR8799_2021}; \citealt{Wang_Ji_HR8799_c_2022}). 
\par In addition, the potential pathway of formation to observe a multi-companion directly imaged system in its present configuration has also been extensively studied (e.g., \citealt{2011ApJ...729..128C}).  For example, in the HR 8799 system, the composition of the planets is largely the same due to sharing a protoplanetary disk, and based on their large separations, HR 8799 bcde represent an extreme scenario of one of two leading formation theories. Either they are at the low mass end of the distribution of companions formed by a gravitationally unstable protoplanetary disk (e.g., \citealt{Kratter_grav_instab_2016}) or they are planets formed via the pebble accretion pathway (e.g., \citealt{Helled_2014prpl.conf..643H,pebbleaccretion_lambrechts2012,lambrechts_coresofgiantplanets_2014}) and represent the farthest separations from the host star enabled via pebble accretion \citep{2011ApJ...729..128C}. Indicators like the carbon-to-oxygen ratios and metallicity of planets have been used to try to understand the system's formation pathway(s), highlighting the importance of the composition of the protoplanetary disk and level of solid accretion during planetary formation \citep{2023arXiv231000088W}. 

\subsection{Multi-companion system HD 206893}
\par Another exemplary case for a multi-companion system is the HD 206893 system that, at present, stands alone as it harbors both a brown dwarf-mass and planetary mass companion: a $26 $\plusminus{3.7}{3.6} $\text{M}_{\text{Jup}}$ outer companion, HD 206893 B, located inside a circumstellar debris disk, as well as a planetary-mass companion, HD 206893 c at $11.5$\plusminus{2.4}{2.2} $\text{M}_{\text{Jup}}$ \citep{Hinkley_2023_c}.  The presence of brown-dwarf mass companions in multi-companion systems presents an interesting test to the planetary formation theories of gravitational instability  and pebble accretion . The coexistence of HD 206893 B and c at close separations ($\leq$10 au) implies that pebble accretion or gravitational instability should be able to create planets with masses spanning from the smallest, rocky planets to brown dwarf-mass planets, even if these systems then undergo dynamical changes that cause the companions to migrate \citep{Dawson_hot_J_migration_2018, Dempsey_2021}. Formation location tracers, such as the C/O ratio derived from atmospheric analysis \citep{oberg_icelines}, could prove important to uncovering the history of this massive system. 

\par HD 206893 B has also been identified as an L/T transition object with an extremely red H-L$^\prime$ color \citep{Milli_discovery}. Substellar companions in the L/T transition have proven difficult to model (e.g., \citealt{Brock_2021}), showing the major impact of inhomogeneous clouds as an opacity source in the L/T transition \citep{Marley_2010_clouds, Apai_2013_LT_clouds}. Near-infrared fluxes have been shown to change drastically throughout the L/T transition (effective temperature $\sim 1400$ K$-1600$ K, e.g.,
\citealt{2005ARA&A..43..195K,Dupuy_Liu_LT_browndwarfs_2012}). Adding to the complexity of modeling, both young and old L dwarfs have shown evidence of reddening in the near-infrared due to dust, generally composed of iron (Fe),  enstatite (MgSiO$_{3}$),  and/or forsterite (Mg$_{2}$SiO$_{4}$)  \citep{Brock_2021}. The spectrum of HD 206893 B has proven difficult to model both with self-consistent atmospheric model grids (e.g., \citealt{Ward-Duong_GPI_spec}) and free retrievals (e.g., \citealt{Kammerer_Gravity_K_band}), with temperature measurements ranging from $\sim1200$K--1800K  and surface gravity measurements ranging from \logg $\simeq$ 2.87--5.0  \citep{Kammerer_Gravity_K_band,Ward-Duong_GPI_spec}, all deriving from low-resolution spectra. These difficulties in modeling point to the need for highly specialized atmospheric models with cloud, dust, and chemistry parameterizations.


\par One method of adding useful information through modeling is to use high-resolution spectra, characterizing individual absorption lines of molecules like CO, H$_{2}$O, CH$_4$, etc., at a resolution of $\geq$ 20,000. \citealt{Xuan_2022} showed that retrieving bulk parameters from high-resolution spectra is largely insensitive to choice of cloud model, even for L-type companions, which is very useful for characterizing the atmospheres of L/T transition objects which are often difficult to constrain with low-resolution spectra or broadband photometry alone \citep{Xuan_2024_survey}. Combining information from low-resolution spectra and photometry, which can provide information about cloud deck location and extinction, with high-resolution spectra, which probes a larger extent in pressure in the atmosphere, are complementary approaches to characterizing exoplanet atmospheres. The Keck Planet Imager and Characterizer (KPIC; \citealt{Mawet_2017_KPIC,Jovanovic_2017_KPIC,2021JATIS...7c5006D}) is a high-resolution spectrograph tailored to high-contrast exoplanet characterization, and has the unique capability of coupling diffraction-limited light from the W. M. Keck Observatory facility Adaptive Optics (AO) system into a single-mode fiber, which is then directed to the NIRSPEC spectrograph \citep{1998SPIE.3354..566M,Martin_2018_NIRSPEC_upgrade} and dispersed at R $\sim$ 35,000. For the case of HD 206893 B, our high-resolution data set allows us to probe both atmospheric composition as well as the orbit (through the addition of orbital radial velocity data). We use the individual resolved absorption lines in the companion's spectrum to infer the bulk properties of the atmosphere, as well as better constrain the orbital plane (even when high precision astrometry already exists).


\subsection{Organization of HD 206893 B Study}
\par In this paper, we present a detection and characterization of HD 206893 B with KPIC at high spectral resolution. In Section \ref{sec:system}, we describe the properties of the system and the findings of previous work, while in Section \ref{sec:observations}, we describe the observations taken with Keck/KPIC. Section \ref{sec:data} details the data reduction steps with the KPIC Data Reduction Pipeline \footnote{https://github.com/kpicteam/kpic\_pipeline}. Section \ref{sec:results} discusses using our forward-model framework to infer bulk parameters of the atmosphere, including effective temperature, gravity, projected spin velocity, measured radial velocity, and carbon-to-oxygen ratio. We then use substellar evolution models to estimate the age, mass, and radius of HD 206893 B. We use the radiative transfer code, \texttt{petitRADTRANS} \citep{Molliere_2019_prt_code}, to perform an atmospheric free retrieval to compare with our forward model analysis. Finally, we present an orbital fit to archival astrometry and radial velocity data including KPIC-measured radial velocity points in two epochs. We then discuss the system's stability based on estimates of HD 206893 c's orbit from \citet{Hinkley_2023_c}. A discussion of the results follows in Section \ref{sec:discussion}. A summary of the work is presented in Section \ref{sec:conclusion}, where we also discuss what future measurements with KPIC and instruments like JWST/NIRSpec may bring. 

\section{System Properties} \label{sec:system}

\par Previous spectroscopic observations of this system have provided broadband photometry from \textit{y-}band to \textit{M-}band (1--4.7 \micron, \citealt{Milli_discovery,Delorme_2017,Grandjean_2019,Meshkat_near_to_thermal,Romero_2021_sphere_disk_non_Detection}), low- to medium-resolution spectroscopy \citep{Ward-Duong_GPI_spec, Kammerer_Gravity_K_band, Hinkley_2023_c} of HD 206893 B, as well as radio imaging of HD 206893 with ALMA \citep{Marino_2020_disk_alma, Nederlander_2021_alma}.

\par HD 206893 is a F5V star with an effective temperature of $\sim$6700 K \citep{Zakhozhay2022starrv} and near-solar metallicity located $40.77 \pm 0.06$ pc away \citep{GaiaDR3_2023}. The star is known to have a debris disk due to the detection of an excess of infrared flux \citep{moor_2006_debris_disk}, which made it a target in the SPHERE High Angular Resolution Debris Disc Survey (SHARDDS, \citealt{wahhaj2016shardds}). The circumstellar environment was imaged in \textit{H}-band with VLT/SPHERE revealing a substellar companion at 270 mas separation \citep{Milli_discovery}. Interestingly, additional photometric measurements in L$^\prime$ \citep{Milli_discovery} and \kband \citep{Delorme_2017} implied an extremely red color, placing the object in the vicinity of an L5--L9 field brown dwarf on a color-magnitude diagram \citep{Ward-Duong_GPI_spec}. \citealt{Delorme_2017} and \citealt{Ward-Duong_GPI_spec} also report a water absorption feature at 1.4 $\mu$m. 
Taken together, these findings have suggested a young, dusty object that proves difficult to fit using current models without the addition of sources of extinction such as high-altitude dust clouds composed of enstatite grains, as detailed by \citet{Kammerer_Gravity_K_band}. A spectroscopic analysis of HD 206893 B at low-resolution (R $\sim$ 500) with VLTI/GRAVITY \citep{GRavity_2017}  in \kband used BT-Settl-CIFIST  \citep{allard_2012_btsettl}, DRIFT-PHOENIX \citep{Helling_2008_DRIFT-PHOENIX}, and Exo-REM \citep{Baudino_2015_exorem}  atmospheric models to infer a temperature of 1049--1600 K and log\textit{g} of 3.50--5.14 \citep{Kammerer_Gravity_K_band}. In addition, \citet{Kammerer_Gravity_K_band} performed a free retrieval on the VLTI/GRAVITY spectrum using both \texttt{petitRADTRANS} \citep{Molliere_2019_prt_code} and \texttt{ATMO} \citep{ATMOcode_Tremblin_2015} codes. The bulk properties retrieved by \prt yielded a lower temperature and lower \logg of 1216\plusminus{13}{17}K and 2.87\plusminus{0.63}{0.47} respectively, while the \texttt{ATMO} retrieval yielded a temperature of 1113\plusminus{51}{52}K and \logg of 2.72\plusminus{0.24}{0.10}.

\par \citet{Delorme_2017} showed that the measured angular separation of  $\sim$270 mas cannot account for the shape of the outer debris disk reported by \citet{Milli_discovery}, and demonstrated that an injected exoplanet of no more than a few \Mj could account for the observed orbit/disk configuration, though no such companion has been detected to date. The debris disk structure was characterized by ALMA and found to begin at a semi-major axis of 30 au from the host star, with a 27 au-wide gap at 74 au from the host, and a full extent to 180 au. Although the present shape of the debris disk was thought to be undisturbed by \planet \xspace \citep{Marino_2020_disk_alma}, recent work has shown that the gravity of HD 206893 B and the self-gravity of a massive debris disk could have had influence in shaping the inner edge of the debris disk, though this result will require higher resolution spatial imaging of the debris disk to confirm that the disk is wider towards the pericenter of \planet's orbit \citep{Sefilian_disk_shaping_2021}. In addition, both of these works do not include the effects of HD 206893 c. 

\par Astrometric data that can be used to map the \planet 's\xspace orbit has been reported since its discovery in 2015. \citet{Grandjean_2019} used imaging data from VLT/SPHERE  \citep{Beuzit_2019_sphere}  and VLT/NACO \citep{2003SPIE.4841..944L,2003SPIE.4839..140R} from 2015--2018 in combination with \text{HIPPARCOS} \citep{Hipparcos} and \textit{Gaia}  \citep{gaia_2016} and radial velocity measurements of HD 206893A from the La Silla/HARPS high-resolution spectrograph \citep{2003Msngr.114...20M}  to further constrain the orbit.  This analysis showed that the presence of an inner companion was needed to explain an observed radial velocity drift and astrometric acceleration, predicted to be $\sim$15 \Mj based upon the size of the drift as well as the detection threshold of the previous observations. Exploration of the orbital architecture of \planet \xspace revealed the presence of an inner companion (HD 206893 c, \citealt{Grandjean_2019, Hinkley_2023_c}).
Using the VLTI/GRAVITY interferometric instrument, \citet{Kammerer_Gravity_K_band} added two high-precision astrometric points and derived a best-fit orbit under two scenarios: an inclined orbit relative to the debris disk and a coplanar orbit with the debris disk.  In particular, the case with an inclined orbit shows a 154\plusminus{12}{9} degree inclination, a more eccentric orbit ($e_\mathrm{incl}$ = 0.29\plusminus{0.06}{0.11} vs $e_\mathrm{copl}$ = 0.13\plusminus{0.05}{0.03}) and a shorter semimajor axis ($a_\mathrm{incl}$ = 9.28\plusminus{1.77}{0.93} au vs. $a_\mathrm{copl}$ = 11.37\plusminus{1.09}{0.75} au).
\par Due to the discovery of the inner companion, the architecture of the system becomes even more complicated, with a mutual inclination between the orbit of HD 206893 B and HD 206893 c of 9.1\plusminus{5.6}{4.9}\degree \citep{Hinkley_2023_c}. \citet{Hinkley_2023_c} also investigate the stability of the system, and find that the massive planets might be in a 9:2 mean motion resonance. In addition, they perform N-body simulations of the system for 10$^5$--10$^6$~yr, noting that the timescale made minimal difference in the results. However,  even $\sim$1 Myr is substantially less than the age of the system, previously estimated to be 155\plusminus{15}{15} Myr \citep{Hinkley_2023_c}.

 \begin{deluxetable}{ccc}
    \tablehead{\colhead{Property} & \colhead{Value} & \colhead{Ref.}}
    \startdata
        & \textbf{HD 206893} &  \\
        \hline
        RA & $21_{\text{h}} 45_{\text{m}} 21_{\text{s}}.91$ &  \citealt{2020yCat.1350....0G} \\
        Dec & $-12\degree 47' 00 06''$ &  \citealt{2020yCat.1350....0G} \\
        Distance & 40.77 $\pm$ 0.06 [pc] & \citealt{GaiaDR3_2023}\\
        Sp. Type & F5V & \citealt{Gray_2006} \\
        $K_{\rm mag}$ & 5.593 $\pm$ 0.021 [mag]& \citealt{2003yCat.2246....0C}  \\
        \Teff & 6680 [K] &\citealt{Zakhozhay2022starrv} \\
        $\logg$ & 4.34 [cm s$^{-2}$ dex] & \citealt{Zakhozhay2022starrv}\\
        $\rm [Fe/H]$ & 0.08 & \citealt{Zakhozhay2022starrv}\\
        \hline
         & \textbf{HD 206893 B} & \\
        \hline
        Period &  25.6 $\pm$1.2 [yr] & \citealt{Hinkley_2023_c} \\
        Separation & 197.23$\pm$0.82 [mas] & \citealt{Hinkley_2023_c}\\
        Pos. Angle & 351.484 $\pm$0.054 [$\degree$] & \citealt{Hinkley_2023_c} \\
    \enddata
    \caption{Stellar and planetary properties for  HD 206893 A and B.} 
    \label{tab:props}
\end{deluxetable}

\par Two of the main aims of these works have been to investigate the orbital configuration of the system as well as explain the ubiquitous reddening seen in the companion's spectrum.  Low-resolution spectroscopy confirmed \planet \xspace as the reddest known substellar object with a ($J_{MKO} - K_{MKO}$) color of $3.36 \pm 0.1$ mag \citep{Ward-Duong_GPI_spec} (compared to the reddest know field L/T transition field brown dwarf, VHS J183135.58--551355.9, with ($J - K_{s}$) of $3.633 \pm 0.277$ mag, \citealt{2024AJ....168...66B}) and required the use of additional dust extinction to be included in standard model grids to reproduce the observed fluxes in \textit{J-}, \textit{H-}, and \kband.


\section{Observations} \label{sec:observations}

\subsection{Instrument Setup}

\par KPIC leverages both the Keck-II 10m-class telescope and its adaptive optics (AO) system as well as the inherent stability of a single-mode fiber-fed spectrograph. The large telescope aperture and atmospheric correction allow it to directly detect companions like \planet \xspace at high spectral resolution (R $\sim 35, 000$). The use of single mode fibers in KPIC provides a stable line spread function and some starlight suppression without the use of a traditional coronagraph due to the single-mode fiber's small numerical aperture and acceptance angle which tends to reject ``off-axis" light from the host star which is separated spatially on-sky. To set up NIRSPEC, we used a custom pupil stop designed for use with KPIC to minimize the thermal background of the instrument during science observations. These hardware upgrades installed for use in KPIC are specified in \citealt{2021JATIS...7c5006D}.  The NIRSPEC filter wheels were set to ``Thin" and ``Nirspec-7" for further background suppression. In 2022, an upgrade mission was performed to add capability to KPIC as well as install a new \kband filter in NIRSPEC \citep{Echeverri_SPIE_2022}, which was used in the observation of \planet \xspace in July 2022. These improvements resulted in an average of 1.5x the throughput of KPIC Phase I. We used the 0.0679 $\farcs$ x 1.13 $\farcs$ NIRSPEC slit, along with the echelle grating rotated to 63\degree\ and cross-disperser at 35.76\degree\xspace to sample wavelengths from 1.94 to 2.41 \micron, which includes both the CO bandhead absorption feature at $\sim$2.3 $\mu$m and H$_2$O features in \kband.

\par We took dark frames of the detector at all required read times after observations concluded to subtract the thermal background of the instrument as well as identify the bad pixels on the detector to mask them from science frames. 

\begin{deluxetable*}{ccccccc}
\tablehead{\colhead{Data} & \colhead{Target} & \colhead{Int. Time [s]} & \colhead{Exposures} & \colhead{Purpose} & \colhead{Fibers Used} & \colhead{Peak Throughput}}
    \startdata
    \hline
    2020-09-29 & HIP 95771 & 1.5 & 5 Exp. x 4 Fib. & Wavelength Solution & 1, 2, 3, 4 & 2.1\% \\
    2020-09-29 & $\lambda$ Cap & 30 & 3 Exp x 4 Fib. & Trace-finding & 1, 2, 3, 4 & 1.4\% \\
    2020-09-29 & HD 206893 & 30 & 6 Exp x 1 Fib. & Stellar Speckle Modeling & 2 & 2.3\%\\
    2020-09-29 & HD 206893 B & 600 & 9 Exp x 1 Fib. & Science Target & 2 \\
    \hline
    2022-07-20 & HIP 81497 & 1.5 & 3 Exp. x 4 Fib. & Trace-finding and wavelength solution & 1, 2, 3, 4 & 2.4\%\\
    2022-07-20 & HD 206893 & 30 & 6 Exp. x 2 Fib.& Stellar Speckle Modeling & 2, 3 & 2.5\%\\
    2022-07-20 & HD 206893 B & 600 & 6 Exp. x 2 Fib& Science Target & 2, 3 \\
\enddata
    \caption{HD 206893 B Observations}
    \label{tab:obs}
\end{deluxetable*}

\subsection{HD 206893 B Observations}

\par The observing log for HD 206893 B is shown in Table \ref{tab:obs}. We observed HD 206893 B during two epochs, one in August 2020 and one in July 2022.  The host star was used as the natural guide star for AO correction in both epochs.  In 2020, KPIC utilized the Keck AO system with the $R$-band Shack-Hartmann wavefront sensor \citep{Wizinowich_2000_SHWFS}, while in 2022, we used the NIR pyramid wavefront sensor for our observations \citep{Bond_2020_pywfs}.  Our observing strategy for both epochs follows a typical KPIC operation procedure: we began by observing the host star by injecting its light into the science fiber(s), with NIRSPEC exposures lasting 60s. During the first epoch, following astrometric information from the orbital prediction site \texttt{whereisttheplanet} \citep{whereistheplanet} based on data from \citealt{Milli_discovery}, \citealt{Delorme_2017}, and \citealt{Grandjean_2019}, we offset Science Fiber 2 (SF2) to the predicted location of the companion using the KPIC Fiber Injection Unit (FIU). SF2 was chosen to observe the companion because it was determined to have the highest throughput of all science fibers during the daytime calibration of the instrument \citep{Finnerty_SPIE_2022}. We then took 600s exposures with NIRSPEC, which were short enough to allow for unexpected technical issues (e.g., AO control loops opening or intermittent cloud cover) while minimizing the effect of read noise compared to other sources of noise. After approximately one hour of observing HD 206893 B, we moved the science fibers back to the host star to take additional science frames for modeling stellar flux in our data. During the second epoch, we moved both Science Fiber 2 and 3 to the location of the planet and ``bounced" between the two fibers in an ABAB pattern to minimize sky background and eliminate any variability in dark frames due to time-variable fringing seen in previous observations with KPIC (See Section \ref{sec:data}). This fiber bounce pattern is susceptible to changes in the thermal background level on the order the exposure time, though this effect is expected to be small and we see no improvement in background subtraction or nodding overhead by using other patterns, i.e. ABBA. These two science fibers were selected from daytime calibrations identifying them as the highest throughput fibers.
\par In 2022, we employed a slightly modified observing strategy in an attempt to mitigate the effect of a free-space etalon in the optical beam path observed in both KPIC Phase I and Phase II data \citep{Finnerty_KPIC_fringing_2022}. This involved first setting the NIRSPEC instrument rotator to 0$\degree$ and doing our standard fiber offset to center the Science Fiber on the location of the target. After taking four exposures, two in each fiber, we then command the NIRSPEC rotator to flip to 180$\degree$ and simultaneously invert the position angle of the FIU by 180$\degree$. This motion changes the angle of incidence of the incoming light onto the KPIC tracking camera dichroic, which was identified as a potential source of the etalon behavior that we discovered in KPIC data. Changing the angle of incidence was thought to change the periodicity of the fringe pattern caused by the etalon, making it easier to identify and subtract. We took four frames in this instrument configuration before returning to take spectra of the host star. Unfortunately in the analysis of this data, we did not get a strong enough detection of \planet \xspace in the first 4 exposures alone (due to seeing conditions) to identify the fringing periodicity as the Science Fiber used to take data also introduced slight variability \citep{Horstman_SPIE_KPIC_fringing}. Therefore, we cannot draw conclusions regarding the potency of this observing technique for future KPIC observations. Individual datasets that yielded companion detections were individually analyzed and we confirmed that the fringing seen in our dataset did not effect the results of our analysis in Section \ref{sec:mcmcresults}.

\subsection{On-Sky Calibration and \texttt{PHOENIX} Stellar Atmospheric Models}

\par Understanding what affects the raw data from the instrument is critical to the characterization of the companion's atmosphere. With KPIC, we observed a wavelength calibration giant star characterized by having many deep and narrow stellar spectral lines as well as a large, well-known radial velocity value. Naturally, these observations also contain telluric lines, which are also included in our wavelength calibration model. We observed HIP 95771 (RV = $-$85.391 km s$^{-1}$, M0.5 III, \citealt{Soubiran2018_GAIA}) in 2020 and HIP 81497 (RV = $-$48.049 km s$^{-1}$, M2.5 III, \citealt{2020yCat.1350....0G}) in 2022 to calculate a wavelength solution for our instrument setup as well as the featureless A-type star $\lambda$ Cap for telluric calibration and trace finding on the detector in 2020. HIP 95771 was chosen as a wavelength calibrator due to its well-known and stable \textit{Gaia} RV measurement while also having distinct stellar features that could be used as wavelength identifiers in each order of the extracted data. We modeled the stellar spectra using a \texttt{PHOENIX} model \citep{Husser2013_PHOENIX} with solar metallicity, $\text{T}_{eff} = 4000$K and $\text{log} g $ = 1.50 cm s$^{-2}$ dex. The \texttt{PHOENIX} model grid incorporates species like VO, TiO, and H$_2$O in the stellar atmospheric models which provide much better fits to lower mass stars where these species become important opacity sources. The models are calculated assuming hydrostatic and radiative-convective equilibrium and are validated against observational data \citep{Husser2013_PHOENIX}. Similarly, HIP 81497 has a very stable and well-known radial velocity \citep{HIP_Star_Classification, Soubiran2018_GAIA}.

\section{Data Analysis} \label{sec:data}

\subsection{Raw Data Reduction}

\par While this section will provide an overview of the raw data reduction process, for the most comprehensive review of raw data reduction with the KPIC Data Reduction Pipeline (DRP), we point the reader to Section 3 of \citealt{WANG_2021AJ....162..148W}. Here we briefly describe the steps taken in the DRP.

\par At either the completion of daytime instrument calibrations or the end of nightly science observations, dark frames are collected to analyze the background of the instrument. The dark frames are co-added based on the exposure time and bad pixels are identified. The co-added frames are saved for dark subtraction and the locations of the bad pixels are stored for masking from science frames. Background subtraction, however, is not ideal in data reduction because the front optics of Keck AO and KPIC are at room temperature, which fluctuates throughout instrument calibration and science collection. In addition, the throughput between fibers is variable with respect to the properties of the optical fibers, seeing conditions, and AO performance, so we cannot use a different off-axis fiber to model the background effectively. In 2020, only SF2 was used to collect data, so any changes in the background flux are folded into the model of the continuum. In 2022, multiple science fibers were used to nod between frames, effectively taking background and science data simultaneously.

\par The data reduction pipeline then records the positions of the fiber traces on each order of the NIRSPEC detector for our chosen cross-disperser and echelle angle. We find the fiber traces on the detector in each order by observing a bright star, generally the same star as is used to compute the telluric calibration, and fitting a one-dimensional Gaussian across the trace of each fiber in the spatial axis. The center position (pixel location) and standard deviation of each Gaussian are saved for each pixel across the wavelength axis of the detector. 

\par We use optimal extraction \citep{Horne_1986_optimal_Extraction} to measure flux in each fiber in a science frame along the wavelength axis. 1D spectra are saved as the total flux integrated across the Gaussian (the center found from trace finding) in each column of the detector. The uncertainty in the flux is determined from the uncertainty in optimal extraction. The 1D spectra are then wavelength calibrated. Telluric modeling was done using satellite data retrieved by Planetary Spectrum Generator (PSG, \citealt{Villanueva2018_PSG}) at the latitude, longitude, and elevation of Keck Observatory on Mauna Kea. The instrument blaze function is wavelength dependent and was modeled using a 5-node spline fit in each order. The final wavelength solution is then calculated using \ \texttt{scipy.optimize.minimize} when jointly fitting for the stellar spectrum, telluric model, and instrument transmission.

\subsection{Forward Modeling the KPIC Spectrum}\label{subsec:fm}

\par With a KPIC Science Fiber centered on the target of interest, we inject both light from the companion as well as flux from the speckle field of the host star into the fiber traces of NIRSPEC orders. In addition, we also incur telluric absorption as the light traverses Earth's atmosphere and any instrumental effects such as free-space etalon behavior \citep{Hsu_NIRSPECFringing_2021, Finnerty_KPIC_fringing_2022, Xuan_2024_HIP555} and chromatic effects. After extracting the flux from each science fiber and calibrating the wavelength of the spectrum, we analyze our data using the \texttt{breads} forward modeling framework. Based on the principles from \citealt{Ruffio_2019_OSIRIS_MRS_HR8799}, \texttt{breads} jointly models two components that contribute to the extracted flux from raw data frames: the flux contributed from diffracted starlight, and flux from the companion. 

\par One drawback of working with fiber-fed instruments at high resolution is the chromaticity of the diffracted speckle pattern that is coupled into the single-mode fiber. Though small, this chromaticity causes stellar speckles to vary in intensity across the wavelength axis, meaning the companion's continuum is difficult to separate from the spline model of  the full continuum of speckle light with planet light \textit{a priori}. \citealt{Wang_2021_KPIC_SPIE} showed that by using a fourth-order polynomial in each order of KPIC spectra, this chromaticity is captured and allows for a stellar model to be produced without normalizing the continuum. \breads  \ leverages this to model the continuum with a 5-node linear spline interpolation in each order analyzed. This spline model is determined analytically and can then be marginalized over as a nuisance parameter without the need for a posterior sampling via MCMC or equivalent \citep{Ruffio_2019_OSIRIS_MRS_HR8799}. 
\par
\begin{figure*}[h]
    \centering
    \includegraphics[width=15cm]{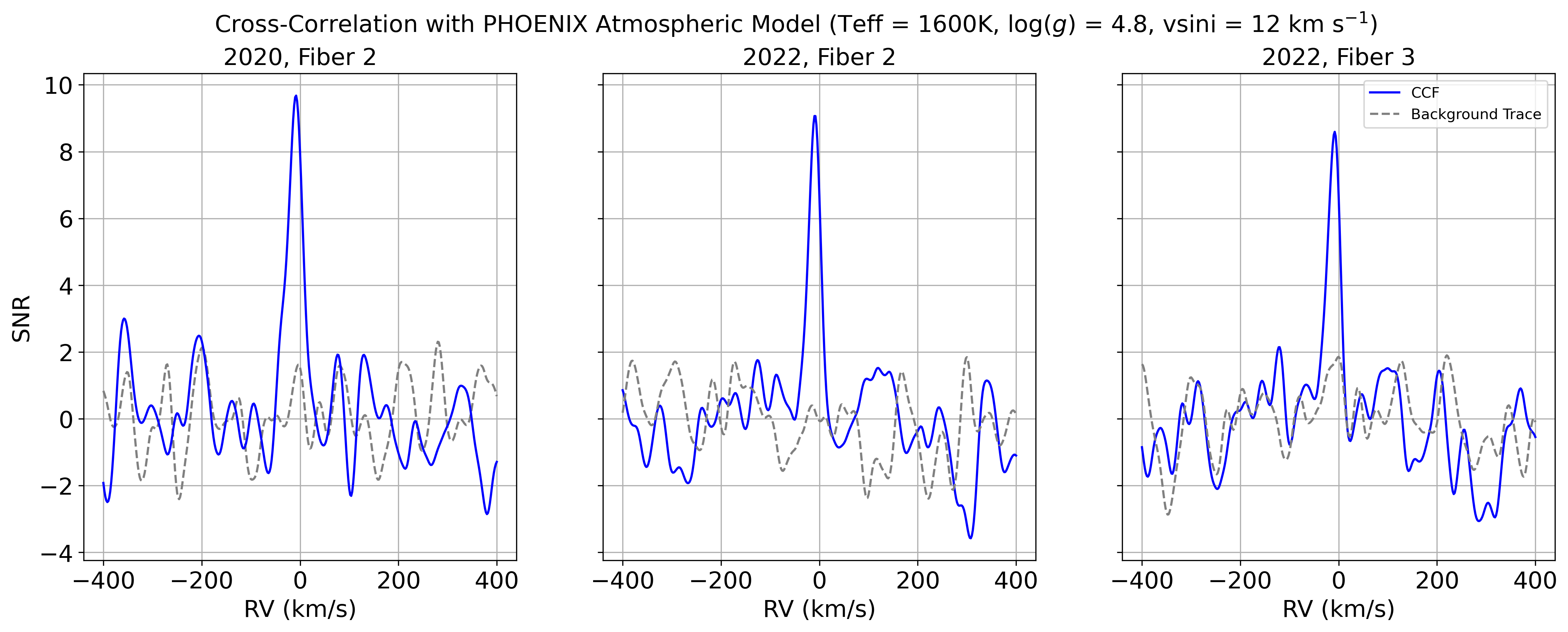}
    \caption{Cross-correlation functions (CCFs) with the custom  \texttt{PHOENIX} atmospheric model using the methodology described in \S \ref{subsec:fm}. The model is selected from a linear interpolation of a fixed model grid described in \S \ref{subsec:customgrid} and has an effective temperature of 1600K with a surface gravity of \logg = 4.8. The model is then spin-broadened to the highest likelihood value, 12 km s$^{-1}$ incorporating both instrumental broadening, expected to be $\sim$ 8 km s$^{-1}$, and projected rotational velocity of the companion. This value was determined by a coarse grid search varying the spin broadening from 0--50 km s$^{-1}$ by 2 km s$^{-1}$.  The solid blue line is the cross-correlation function of the model template with the data on the planet. The dashed grey line is the cross-correlation function of the same template with extracted spectra of a background trace on the detector. The value of the signal-to-noise ratio of the peak of the CCF is scaled based on the wings of the CCF where the average SNR$ = 1$, deriving from formalism of the likelihood model \citep{Ruffio_2019_OSIRIS_MRS_HR8799}.Each individual data set shows the detection of the planet signature at $>8 \sigma$.}
    \label{fig:ccf}
\end{figure*}

 The forward model is defined as,
\begin{equation}
    \textbf{\textit{d}}(\lambda,f) = \textbf{M}_{RV,v\textit{sin i},...} \textbf{$\phi$} + \textbf{n}, 
\end{equation}
where $\textbf{\textit{d}}$ is the data vector as a function of wavelength and fiber number, \textbf{\textit{n}} is a uncorrelated Gaussian noise vector, and \textbf{M} is a matrix of  atmospheric models (e.g., \texttt{BT-Settl, PHOENIX}) that vary with respect to a set of physical parameters such as the radial velocity, spin, and effective temperature. The set of linear parameters of the forward model, \textbf{$\phi$},\ defines the amplitude of the companion flux and stellar flux at each wavelength value. Linear parameters are analytically marginalized to simplify the atmospheric characterization by varying only physical parameters. These physical parameters are referred to as non-linear parameters as varying these only change the definition of the atmospheric model.  Consequently, the likelihood function, $\mathcal{L}(\psi,\phi,s^{2})$ is written as a multivariate Gaussian distribution as described in Appendix D of \citet{Ruffio_2019_OSIRIS_MRS_HR8799}:

\begin{equation}
\mathcal{L}(\psi,\phi,s^{2}) =  \frac{1}{\sqrt{(2\pi)^{N_{d}}|\Sigma_{0}|s^{(2*N_{d})}}} \hspace{1em}\text{exp}\Bigl\{-\frac{\chi^2}{2s^{2}} \Bigr\},
\label{eq:likelihood}
\end{equation}

where $\chi^{2}$ is defined by: 

\begin{equation}
    \chi^{2} = (\textbf{d} - \textbf{M}_{RV,v\textit{sin i},...}\mathbf{\phi})^{\top} \mathbf{\Sigma}_{0}^{-1}(\textbf{d} - \textbf{M}_{RV,v\textit{sin i},...}\mathbf{\phi}), 
\end{equation}
and $N_{d}$ is the number of dimensions, $\Sigma_0$ is the normalized covariance matrix, and $s^2$ is a scaling parameter. For the purposes of confirming a detection of the companion object, the likelihood is maximized using a linear least squares algorithm, varying the RV parameter from $-$400 km s$^{-1}$ to 400 km s$^{-1}$ in steps of 2 km s$^{-1}$ where the best guess of surface gravity and effective temperature along a coarse grid suffice to produce a detection via cross-correlation of the full atmospheric model \citep{Shubh_breads_2023,ruffio_Breads_7672b_2023,Horstman_2024_GQLUPB}. Using this methodology, with a T$_{eff} = 1600$ K, \logg = 4.8 \texttt{PHOENIX} atmospheric model template, we detect \planet \xspace with a SNR of 9.8 in our 2020 data set and an SNR of 12.7 in 2022 when combining our Fiber 2 and Fiber 3 data sets, each showing a detection of $\sim 9\sigma$ with consistent radial velocity measurements. The SNR maps from cross-correlation are shown in Figure \ref{fig:ccf}. 

\par
\begin{figure*}[h]
    \centering
    \includegraphics[width=15cm]{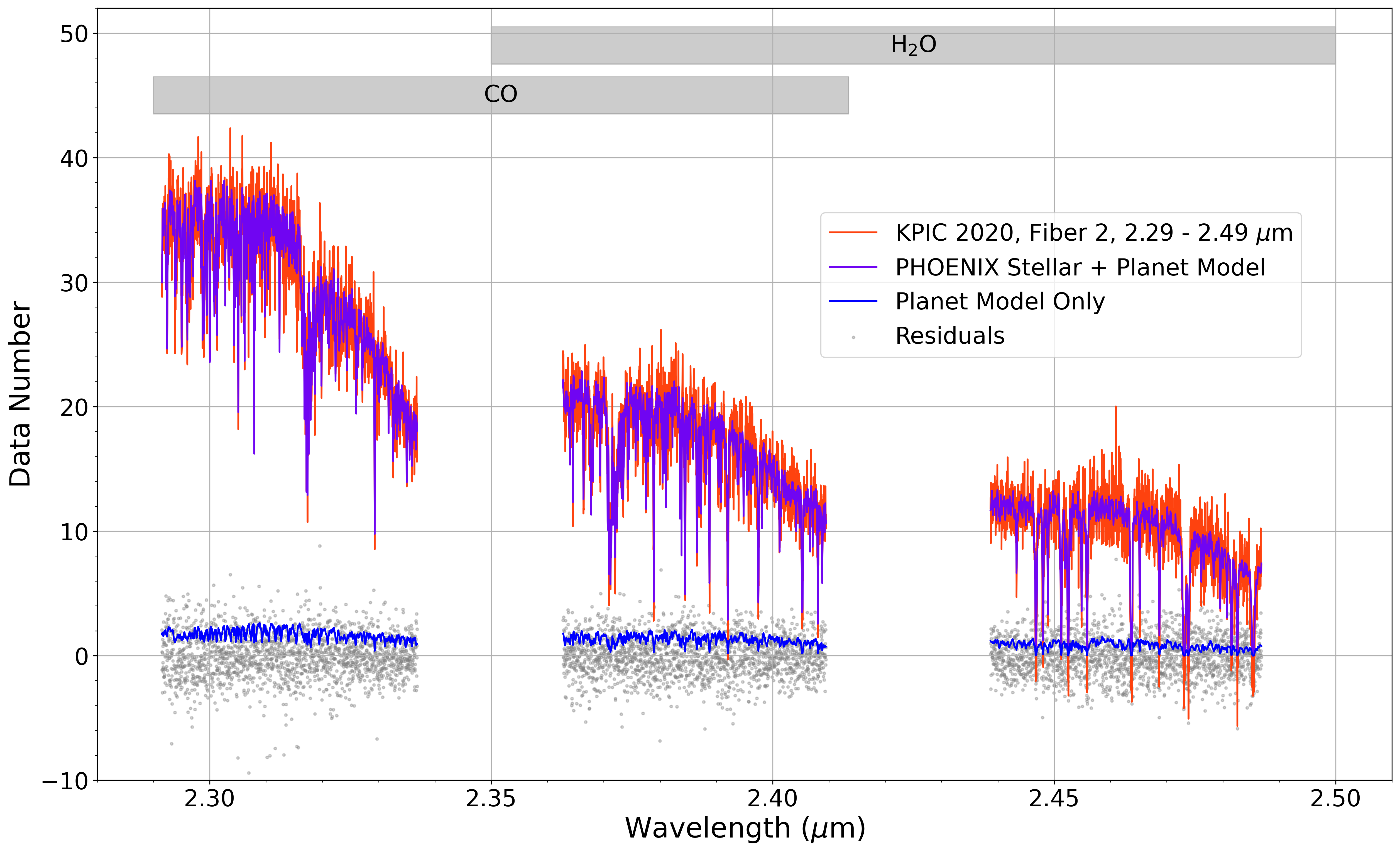}
    \caption{The 1D extracted spectra (red) from KPIC, Fiber 2, with the fiber located at the position of the planet. The best fit \texttt{PHOENIX} stellar plus planet model is shown in purple, with the best-fit planet model plotted separately in blue. The best fit star + planet model is built from a linear combination of the stellar spectrum, taken from our exposures of the host star alone, and the planet model described in Section \ref{subsec:customgrid}. The residual values between the data and the star + planet model are shown in grey points towards the bottom. }
    \label{fig:spectrum_2020}
\end{figure*}

 In this work, we use orders 31--33 of NIRSPEC, the last three of the nine available to KPIC, in \kband which range from 2.29--2.34\,$\mu$m, 2.36--2.41\,$\mu$m, and 2.44--2.49\,$\mu$m.  These orders include the first overtone of the CO ro-vibrational mode at $\sim 2.3\,\mu$m, where the evenly spaced absorption features provide strong evidence of detection from our cross-correlation maps.  An extracted spectrum from our 2020 dataset is plotted in Figure \ref{fig:spectrum_2020}. We generally do not use the first three orders of NIRSPEC, 37--39 (1.94--2.05 \micron), due to the presence of CO$_2$ telluric absorption that varies over time. The three middle orders, 34--36 (2.10--2.22 \micron), lack stellar absorption features for wavelength calibration and are often unreliably calculated and do not affect the SNR of the companion detection. 


\section{Spectral Fitting with Self-Consistent Models}\label{sec:results}

\par Using the \breads \xspace forward-model framework and likelihood defined in Equation \ref{eq:likelihood}, we initiate a Markov Chain Monte Carlo (MCMC) with the python package \texttt{emcee} \citep{emcee_2013} to determine the posterior distribution of parameters varied in the grid, as well as the rotational and radial velocities. The MCMC uses 512 walker with a 10,000 step burn-in and a subsequent 10,000 step sampling. The walkers are initiated at a randomly distributed position in a uniform prior space: from 1200 K--2000 K in effective temperature, 3.5--5.0 in log surface gravity, 0--50 km s$^{-1}$ in \vsini, and $-$50--50 km s$^{-1}$ in radial velocity. At each step, the MCMC computes the likelihood of the regularly interpolated model grid evaluated at the walker values for each data set (2020 Fiber 2, 2022 Fiber 2, and 2022 Fiber 3). We then add the log-likelihoods as opposed to adding the raw data numbers of each dataset. The convergence of the MCMC run is visually determined by examining the walker trace plots, looking for a steady distribution about the mode. 

\subsection{BT-Settl Phoenix Model} \label{subsec:btsettl}

\par An initial analysis of the properties of the planet was performed using the publicly available \texttt{BT-Settl} atmosphere model grids \citep{allard_2012_btsettl}.  The 20,000 step (10,000 step burn-in, 10,000 step sampling) MCMC results with this grid yields bimodal posterior distribution in \vsini and RV, with the corresponding corner plot show in Figure \ref{fig:BTsettlcornerplot}. Upon analyzing the model spectra, it became apparent that these models do not have a smooth transition across the L/T-transition at approximately 1400--1600 K. The MCMC chains revealed that walkers were traversing this L/T transition region (see Figure \ref{fig:btsettlwalkertrace}), but the interpolation between the models meant that our posterior distributions in temperature and \logg are not symmetric with large tails in Figure \ref{fig:BTsettlcornerplot}. We postulate that in this case the models' transition from a cloudy atmosphere to a clear atmosphere at lower temperatures is not a good match to the data \citep{allard_2012_btsettl}. \citealt{Sanghi_2023}  found that BT-Settl models over-predict dust opacity in its \~ 1800 K model, meaning these L/T transition objects found to pile-up in particular temperature ranges like 1500--1600 K (see \citealt{Sanghi_2023} , Section 8.3.1 and Fig. 22).

\subsection{Custom Atmospheric Grid}\label{subsec:customgrid}

\par Due to the known cloud/dust content of the extremely red \planet, we choose to use a custom \texttt{PHOENIX} model grid that was created following the prescription of \citealt{Brock_2021} and \citealt{Barman_2011} which is specifically tailored to better model L/T transition atmospheres. This model grid has two variable parameters: effective temperature and surface gravity, which range from $T_\mathrm{eff} = 1200$--2000 Kelvin to $\text{log}(g) = 3.5$--5.0, respectively. The wavelength axis ranges from 1.7--2.9\,$\mu\text{m}$ with 0.01\,$\angstrom$ sampling. The \texttt{PHOENIX} model atmospheric code is a 1-D, self-consistent radiative transfer model that produces synthetic spectra line-by-line using up-to-date molecular line opacities \citep{Tennyson_Yurchenko_2012_exomol}. These models are made under the assumptions of hydrostatic, chemical, and  radiative-convective equilibrium. The models also assume solar metallicity and a C/O ratio of 0.55, though we vary this parameter in a subsequent model grid detailed in Section \ref{subsec:co}. Previous work and photometric data point towards \planet \xspace being an L/T transition object with excess amounts of extinction due to dust in the \textit{J}- and \textit{H}-band \citep{Ward-Duong_GPI_spec}. Testing cloud extents, properties, and other parameterizations has become increasingly important as these L/T transition objects are discovered and characterized as they deviate quite strongly from earlier generation atmospheric modeling codes \citep{Brock_2021}. To account for the red extinction of \planet, we tuned our model to contain clouds with a mean particle size, $a_{0}$, of 1\,$\mu$m in a log-normal distribution (informed by \citealt{Marley_1999}), and a fixed cloud height parameter (PGS) of $10^6\,\text{dyne cm}^{-2}$ after comparing to a similar model grid set that showed that the lowest pressure enstatite cloud deck created the most \jband and \hband extinction in the atmosphere \citep{Hoch_2022}. In addition, it has been widely suggested that the redness of substellar brown dwarf and planetary mass companions in the L/T transition are best explained by atmospheres out of equilibrium \citep{Marley_Saumon_Cushing_Ackerman_Fortney_Freedman_2012,Tremblin_atmosphere_disequilibrium_2019}, and therefore necessitate a model atmosphere in which disequilibrium chemistry is included. Here we employ this disequilibrium chemistry by driving the vertical mixing of CO in atmospheric layers, tuning the coefficient of eddy diffusion, $K_{zz}$. In this model grid, $K_{zz} = 10^{8}$ and is kept constant, informed by prior modeling of L/T transition brown dwarfs by \citet{Brock_2021}.

\subsection{KPIC High-Resolution Spectrum + Forward Model MCMC Results with Custom \texttt{PHOENIX} Grid} \label{sec:mcmcresults}
 
\par
\begin{figure*}[!h]
    \centering
    \includegraphics[width=0.75\textwidth]{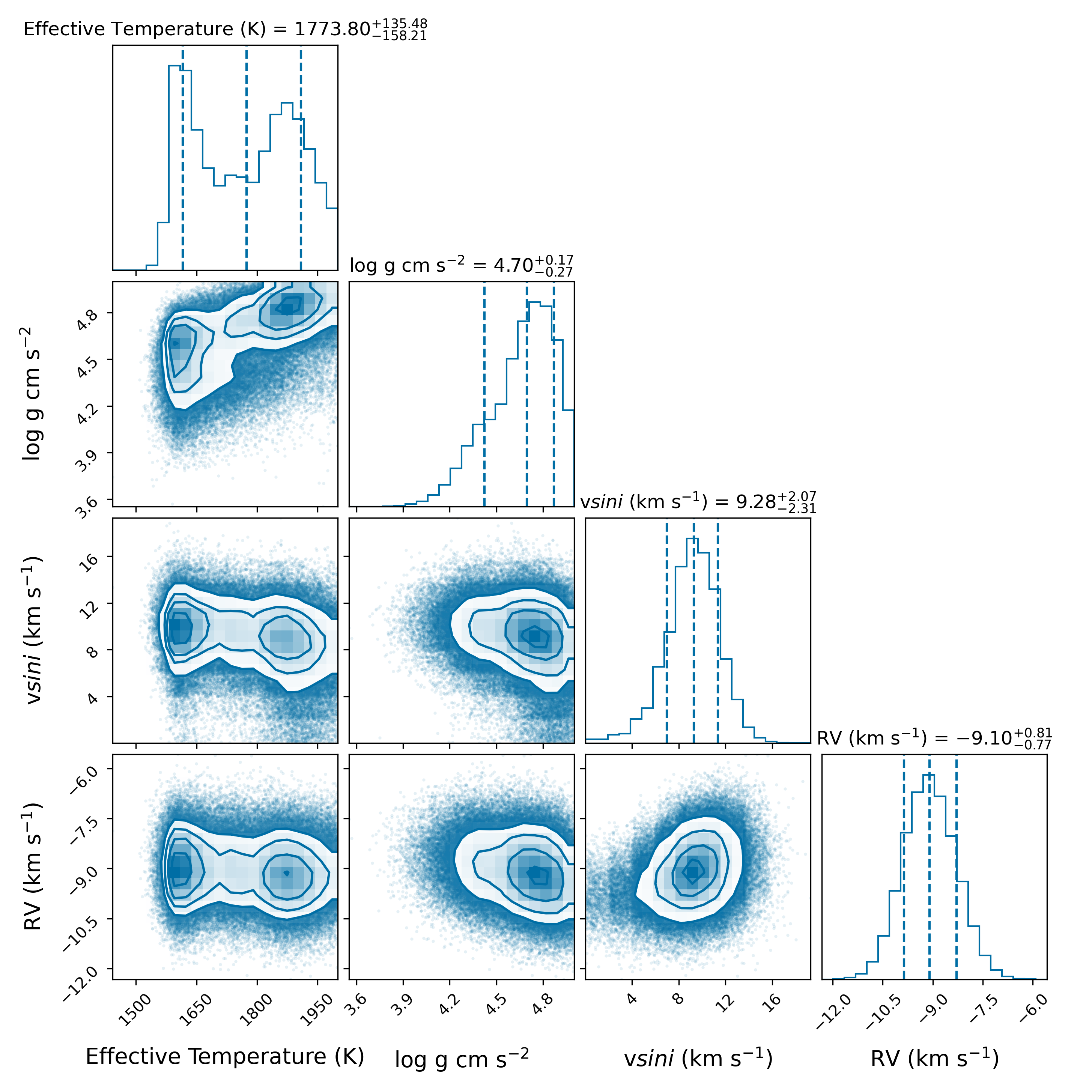}
    \caption{Corner plot for HD 206893 B showing effective temperature, surface gravity, \vsini, and RV using the full \texttt{PHOENIX} model grid and the KPIC \kband spectra.}
    \label{fig:fullgrid_atmo}
\end{figure*}

 Using the same forward model framework and MCMC setup, we ingest the custom \texttt{PHOENIX} grid to model the atmosphere of \planet. The posterior distributions of the MCMC are shown in Figure \ref{fig:fullgrid_atmo}. We see a bimodal distribution in effective temperature, with one peak at $\sim$1600 K and the other at $\sim$1900 K. Although the temperature peaks correspond to the same RV and \vsini values, the temperature vs \logg plot points to two solutions:  (1600 K ; \logg = 4.6 cm s$^{-2}$ dex) and (1900 K; \logg = 4.8 cm s$^{-2}$ dex). In a previous analysis of KPIC data of HR 8799, a correlation was found between effective temperature and surface gravity, which may be attributed to a degeneracy in the depth of CO absorption features in models with high temperature/high \logg and low temperature/low \logg (relative to each other) \citep{Wang_HR8799_2021}. Varying \logg in these atmospheric models changes both the shape of the absorption features, especially around the CO band head ($\sim 2.3 \mu$m), as well as the overall continuum level and shape with $\logg = 5.0$ models exhibiting a flatter continuum than the $\logg = 4.5$ model at the same temperature. Coupled with the fact that these model grids are, by default, assuming solar abundances of carbon and oxygen, fitting high-resolution data by eye without considering the implications of the selected abundances of C and O (as well as the C/O ratio) and other physical properties like the companion radius and age will lead to nonphysical solutions of the companion's bulk properties. As a result, our surface gravity measurement, to 68th percentile, lies at 4.70\plusminus{0.17}{0.27} cm s$^{-2}$ dex. We also see evidence of spin broadening in the spectrum, with the measured value at 9.3\plusminus{2.1}{2.3} km s$^{-1}$. Finally, we measure a barycentric radial velocity shift of $-$9.10\plusminus{0.81}{0.77} km s$^{-1}$. The measured radial velocity of the host star HD 206893 A is $-$12.45 $\pm$ 0.59 km s$^{-1}$ \citep{Soubiran2018_GAIA}, which implies a relative RV shift of 3.35 km s$^{-1}$.

\subsection{Low-resolution Spectrum with Gemini Planet Imager Integral Field Spectrograph (IFS)}

\par For a $\Teff\approx1400$ K L/T transition brown dwarf companion, \citet{Xuan_2022} found that the results of fitting $K$-band high-resolution spectra are independent of the chosen cloud model because the cloud bases lie below the $K$ band emission contribution function. However, they showed that low-resolution spectra (LRS) from 1--2.2\,$\mu$m are strongly affected by clouds, specifically at the $\approx1\,\mu$m region where the continuum overlaps with the cloud base of MgSiO$_3$. This shows that broadband LRS tends to be more sensitive to clouds. We aim to use LRS as a validation of our \texttt{PHOENIX} atmospheric model grid and to break the degeneracy seen in our posterior distribution of temperature and log$(g)$ with a physically-motivated explanation.

\par We fit the \texttt{PHOENIX} models to \kband Gemini Planet Imager data of HD 206893 B \citep{Ward-Duong_GPI_spec} using the \texttt{Spectral Modeling Analysis and RV Tool} (\texttt{SMART}) forward modeling code \citep{Hsu_smart_2021ApJS..257...45H}. \texttt{SMART} is a Bayesian spectral fitting code that is able to ingest any model grid and spectrum. Briefly, \texttt{SMART} does an initial MCMC with a burn-in fit to estimate effective temperature, surface gravity, rotational broadening, RV, and instrumental broadening (as well as nuisance parameters for  flux and wavelength calibration). It then compares the best-fit spectrum to the science spectrum and identifies any discrepancies that could be caused by poor cosmic-ray rejection, bad pixels, etc. With this information, the code creates a bad pixel mask and runs a second MCMC with the walkers initialized within a few percent of the best-fit values from the initial run. The choice of prior values can also be limited in the second MCMC, but here we leave them the same in both MCMC runs.  In this case, with a published science spectrum, we elect to mask the outer edges of the \kband bandpass filter, as we see both uncertainties inflate at the shorter and longer ends of \kband, near the atmospheric water bands. 

\par
\begin{figure}
    \centering
    \includegraphics[width=\linewidth]{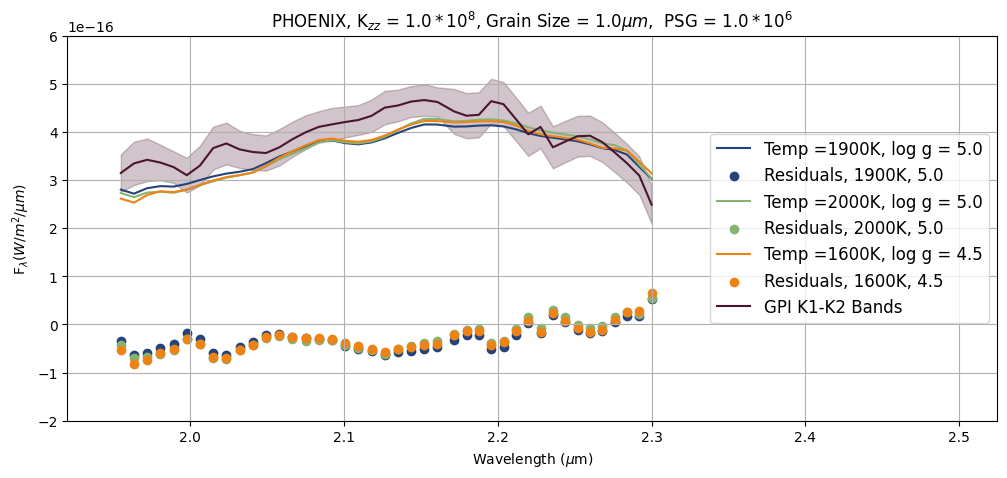}
    \caption{GPI IFS \kband Data (1.95--2.30\,$\mu$m), shown with shaded uncertainties representing the $1-\sigma$ envelope, with the three best fit broadened custom \texttt{PHOENIX} models. All three models fit the data well and are similar to one another despite the 400 K variation and 0.5 dex change between the 2000 K, $\logg = 5.0$ and 1600 K, $\logg = 4.5$ model. However, the higher temperature models imply a smaller companion radius of 0.69 \Rj, whereas the lower temperature, lower log$(g)$ model equates to a more physically-motivated radius of 1.0 \Rj.}
    \label{fig:PHOENIX_gpi_fit}
\end{figure}

 Gemini Planet Imager \citep{Macintosh_gpi_2008} IFS observed \planet \xspace as part of the Gemini Planet Imager Exoplanet Survery (GPIES; \citealt{macintosh_gpi_2014,Nielsen_2019_GPIES}) with the spectra published in \citet{Ward-Duong_GPI_spec}. Using this low-resolution \kband spectrum with our custom \texttt{PHOENIX} model grid (broadened to appropriate resolution) in \texttt{SMART}, we find a preferred temperature of $1730$\plusminus{80}{90} K, and a log$(g)$ of $4.9$\plusminus{0.1}{0.2}. We fit for \kband separately here as the custom \texttt{PHOENIX} model does not extend in wavelength to \jband and \hband. The high temperature, high-gravity model fits the \kband data well except for at the reddest wavelengths where the best-fit model flux is too high. We perform a grid search using the \texttt{PHOENIX} models to visualize the model fit without modulating the spectra to fit for instrument parameters as in \texttt{SMART}. We scale the model flux and resolution to realistic levels for the GPI data and perform a simple $\chi^2$ fitting routine. We find that the best-fit models are T$_{eff}$ = 1900 K, log$(g)$ = 5.0; T$_{eff}$ = 2000 K, log$(g)$ = 5.0, and T$_{eff}$ = 1600 K, log$(g)$ = 4.5, the last of which is consistent to the results from our forward model of the high-resolution KPIC spectra. These select models are shown against the data in Figure \ref{fig:PHOENIX_gpi_fit}. 

\par
\begin{figure*}[!h]
    \centering    \includegraphics[width=0.7\textwidth]{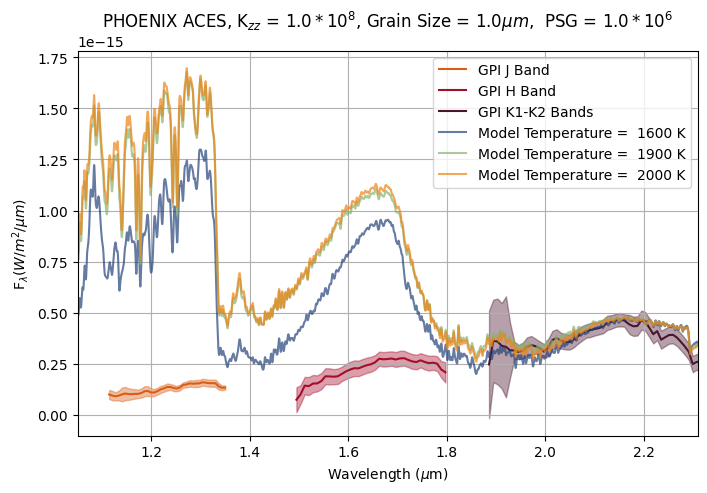}
    \caption{Full GPI IFS Spectrum (\jband: $1.114 - 1.350$ $\mu$m, \hband: $1.495 - 1.797$ $\mu$m, \kband: $1.886 - 2.396$ $\mu$m) of \planet \xspace. The $1-\sigma$ uncertainty envelope is shown in the matching shaded color. The best-fit \texttt{PHOENIX-ACES} \kband models are plotted and scaled to match peak flux in \kband, showing the previously observed extreme extinction in \jband and \hband.}
    \label{fig:PHOENIX_gpi_alldata}
\end{figure*}

 While none of these models seem to match the data exceedingly well at the long end of \kband passed the CO absorption overtone absorption feature at 2.3 $\mu$m, extending the wavelength coverage to the full GPI spectrum shows evidence of extinction in $J-$ and $H-$ bands, a known property of \planet 
 \citep{Ward-Duong_GPI_spec,Kammerer_Gravity_K_band}.  The GPI instrument has a known level of high throughput loss at the end of \kband, which could be the cause of this discrepancy \citep{Maire_2014_spie_gpi}. As the \texttt{PHOENIX} models we use up to this point do not extend to \jband, we rely on the \texttt{PHOENIX-ACES} model set that is based on the same framework and mirror our model grid parameters with respect to particle size, cloud depth, and non-equilibrium chemistry. This custom model set is described in depth in \citet{Hoch_2022}. We match the flux level of each model to the flux maximum in \kband, with the results shown in Figure \ref{fig:PHOENIX_gpi_alldata}.

\par We calculate the implied radius of the companion based on scaling our flux to \kband and find that only the 1600 K/log$(g)$ = 4.5 models produce physically-motivated radii for an object with an age of 155\plusminus{15}{15} Myr and a mass of 26.2\plusminus{3.7}{3.6} \Mj \citep{Hinkley_2023_c} using evolutionary models from \citealt{chabrier_evo_models_2022}. We find a companion radius of R = 1.03\plusminus{0.04}{0.04} R$_{\text{Jup}}$ in \kband whereas with the higher \logg models, the radius of the companion is R = 0.69\plusminus{0.05}{0.05} \Rj.

\subsection{Limited Temperature Priors with \texttt{PHOENIX} Model}\label{sec:tlimit}
\begin{figure*}[!h]
    \centering
    \includegraphics[width=0.6\textwidth]{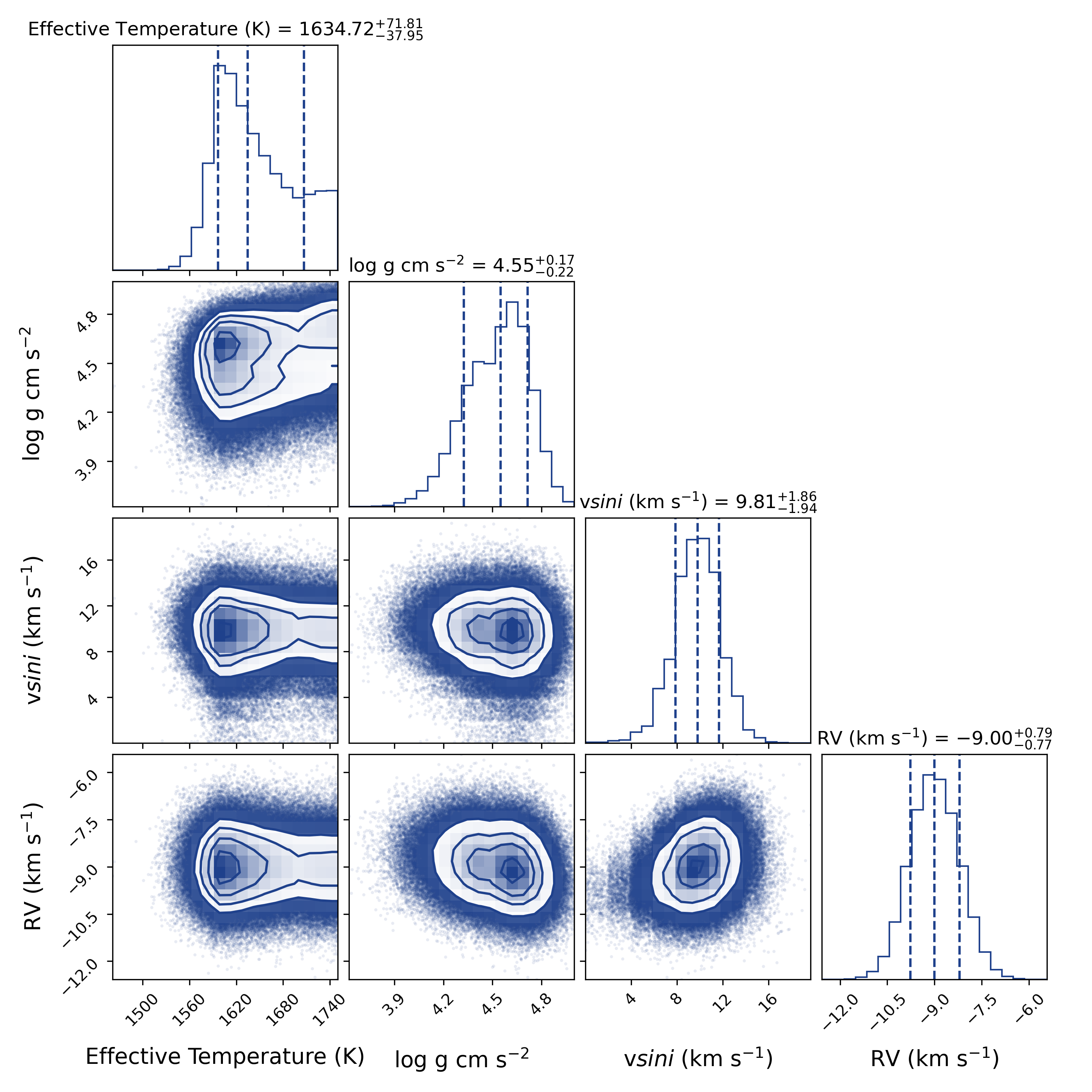}
    \caption{Corner plot for Forward modeling results of KPIC Spectrum of HD 206893 B. We vary the effective temperature, surface gravity, \vsini, and RV. In this corner plot, we limit the prior space from 1200 - 1750 K, motivated by our analysis of the GPI LRS best-fit model and retrieval results.}
    \label{fig:teff_limit_atmo}
\end{figure*}

\par Having ruled out a higher temperature, higher \logg solution due to a nonphysical radius, we elected to limit the effective temperature prior space to a maximum of 1750 K. This value was chosen as it was a local minimum in the posterior distribution of our high-resolution KPIC MCMC. The MCMC configuration remains unchanged; we use 512 walkers with a 10,000 step burn-in and a subsequent 10,000 step sampling. Walkers are initiated at a randomly distributed position in the prior space: from 1200 K - 1750 K in effective temperature, 3.5 - 5.0 in \logg, 0 - 50 km s$^{-1}$ in \vsini, and -50 - 50 km s$^{-1}$ in RV. We vary the \vsini and RV parameter in this MCMC to check for consistency while changing the prior space of the grid. 

\par The results of the MCMC are shown in a corner plot in Figure \ref{fig:teff_limit_atmo}. Limiting the maximum effective temperature eliminates the bimodal behavior we see when using the full temperature grid. The posterior temperature distribution shows a tail towards higher temperatures, but ultimately favors a temperature of 1635\plusminus{72}{38} K. We observe a corresponding decrease in surface gravity from 4.70 cm s$^{-2}$ dex to 4.55 \plusminus{0.17}{0.22}cm s$^{-2}$ dex, which matches our early observation that high-surface-gravity models were paired with high effective temperatures. Both our \vsini value of -9.8\plusminus{1.9}{2.0} km s$^{-1}$ and RV value of -9.00 \plusminus{0.79}{0.77} km s$^{-1}$ are consistent with the full model grid.

\subsection{C/O Measurement} \label{subsec:co}

\par
\begin{figure}[h!]
    \centering
    \includegraphics[width=\linewidth]{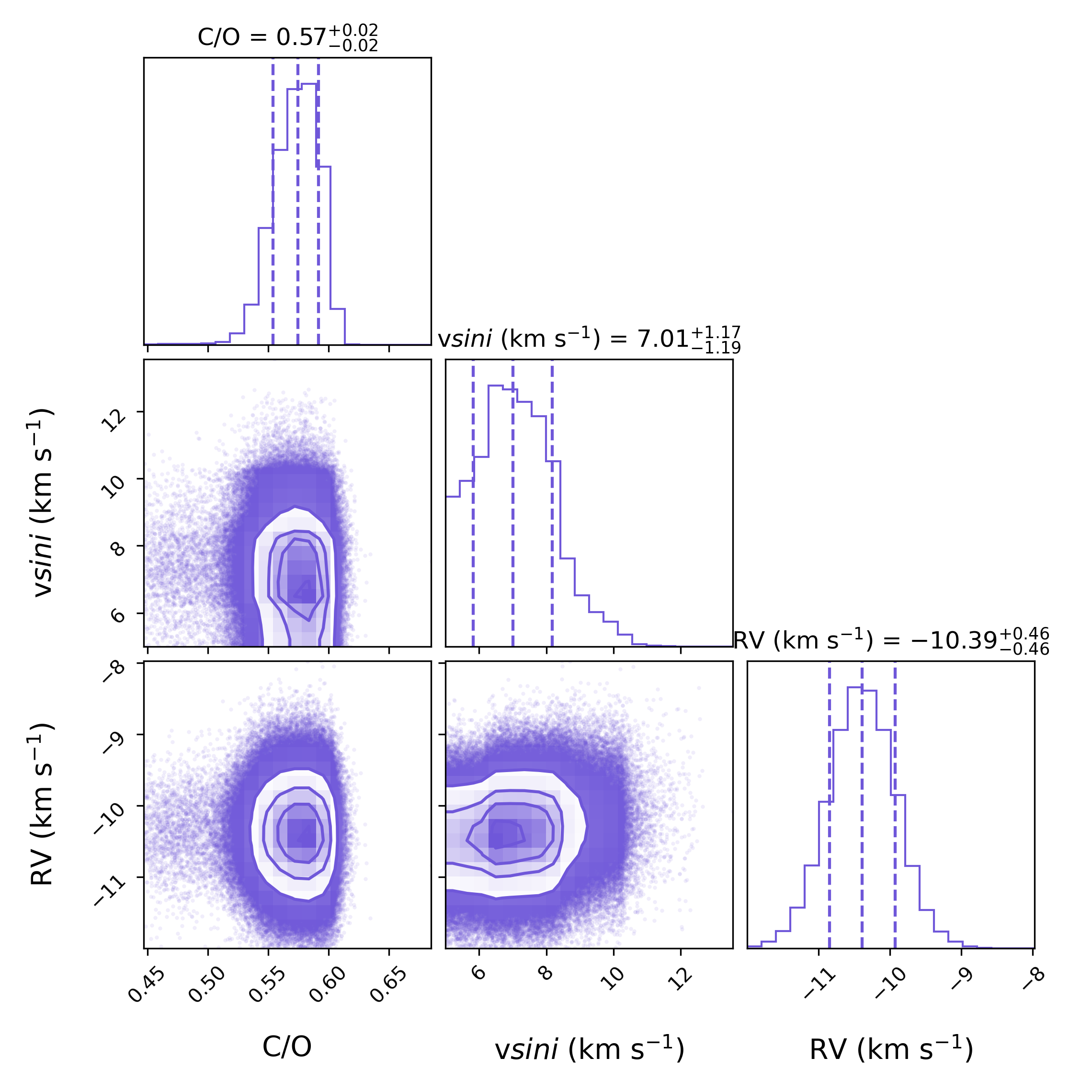}
    \caption{C/O ratio fit using \texttt{PHOENIX} models with T$_{eff}$ = 1600 and \logg = 4.5.}
    \label{fig:coratio}
\end{figure}

 After determining the best-fit atmospheric parameters using the \texttt{PHOENIX} models, we generated a grid with varied carbon and oxygen content at a fixed temperature (1600 K) and \logg = 4.5 cm s$^{-2}$. This model grid was processed in the same matter, interpolated, and ingested into \texttt{breads} where we fit the high resolution KPIC data for the C/O ratio, the radial velocity, and \vsini simultaneously. [C/O] parameter was varied from 0.447 to 0.891 while keeping the overall metallicity solar. The data preferred a value of $0.57 \pm 0.02$, near the solar value, and the corner plot is shown in Figure \ref{fig:coratio}. The fit radial velocity value, $-10.4 \pm 0.5$ km s$^{-1}$ is slightly $>1\sigma$  from the modeled RV from our full atmospheric parameter posterior in \ref{subsec:customgrid} by $\sim$100 m s$^{-1}$, though the companion spin measurement of 9.3 \plusminus{2.1}{2.3} km s$^{-1}$ is consistent with the full atmospheric parameter posterior. 

\section{Spectral Fitting with Retrievals}\label{sec:prt}

\par To verify the results of the forward modeling and determine whether the retrieval can provide additional constraints to the strong atmospheric reddening, we also perform a free retrieval on the GPI \textit{J-,H-, }and \kband spectra simultaneously using the fast radiative transfer code \texttt{petitRADTRANS} \citep{Molliere_2019_prt_code}. We use \prt in its low-resolution correlated-k opacity mode to model the data and retrieve pertinent quantities such as the bulk atmospheric parameters, chemical abundances, and vertical temperature-pressure profile. This method allows for greater flexibility in the model parameter space and has proven to be a useful tool in atmospheric characterization. However, the results of a retrieval must be taken with caution as they do not require the resulting atmospheric model to be fully self-consistent. The parameters varied in this free retrieval are shown with their prior space in Table \ref{tab:prt_params}, in addition to the best fit results. We use \texttt{pyMultiNest} with 4000 live points during the free retrieval, and have 16 free parameters. We fit surface gravity, radius of the planet, the internal temperature, and use a three-node spline temperature model for the vertical structure of the atmosphere akin to the setup of \citealt{2020A&A...640A.131M}. We use the optical depth model suggested by \citealt{Nasedkin_2024} and fit for the parameters $\Delta$ and $\alpha$ in the pressure, P, formulation:
\begin{equation}
    \tau = \Delta\text{P}^\alpha
\end{equation}
where $\Delta$ is a proportionality factor and $\alpha$ is the pressure power law index. In the chemical disequilibrium model, we fit for the carbon quench pressure, log(P$_{\text{quench}}$), i.e. the atmospheric layer where chemical reactions stop occurring due to low pressures, the metallicity, [Fe/H], and the C/O ratio. Finally, we include clouds in our model based upon the Ackerman-Marley cloud model \citep{Ackerman_2001}. This model is parameterized by the particle size distribution, in a log-normal distribution with width $\sigma_{log\_norm}$ [\micron], the vertical mixing parameter, K$_{zz}$ [cm$^2$s$^{-1}$], and the unitless sedimentation parameter, f$_{sed}$ \citep{2020A&A...640A.131M}.

\begin{deluxetable*}{cccc}
    \tablehead{\colhead{Parameter} & \colhead{Best Fit Model (\kband only)} & \colhead{J-, H-, K-band Best Fit Model} & \colhead{Best Fit Model Prior Range}}
    \startdata
    \logg [cm s$^{-2}$ dex]&  3.57 \plusminus{0.74}{0.74}& 2.78 \plusminus{0.22}{0.23}&Uniform(2.0 - 5.5) \\ \hline   
        Radius [R$_{Jup}$]&  0.91 \plusminus{0.18}{0.18}&  2.16 \plusminus{0.02}{0.04}&Uniform(0.9 - 2.2) \\ \hline   
        Effective Temperature [K]& 1640 \plusminus{350}{350}&  1225 \plusminus{}{}&Uniform(500 - 2700) \\ \hline  
        $\alpha$  & 1.454 \plusminus{0.25}{0.25}&  1.367 \plusminus{}{}&Uniform(1 - 2) \\ \hline  
        log($P_{quench}$) [bar]& -1.2 \plusminus{2.5}{2.5}&  -3.4 \plusminus{1.5}{1.4}&Uniform(-6 - 3) \\ \hline 
 log($\Delta$)  & 0.56 \plusminus{0.22}{0.22}&  0.424 \plusminus{}{}&Uniform(0 - 1.0) \\ \hline  
        Fe/H [dex]& -0.13\plusminus{0.72}{0.72}&  -1.39 \plusminus{0.07}{0.06}&Uniform(-1.5 - 1.5) \\ \hline  
        C/O& 0.395 \plusminus{0.19}{0.19}&  0.73 \plusminus{0.07}{0.08}&Uniform(0.1 - 1.6) \\ \hline 
 $\sigma_{log\_norm}$ [\micron]& 1.97 \plusminus{0.58}{0.58}&  1.384 \plusminus{}{}&Uniform(1 - 3)\\\hline  
        log($k_{zz}$) [cm$^2$ s$^{-1}$ ]& 9.71 \plusminus{0.17}{0.17}&  5.92 \plusminus{0.35}{0.25}&Uniform(5 - 13) \\ \hline
        f$_{sed}$& 7.10 \plusminus{0.20}{0.20}&  4.03 \plusminus{0.58}{0.51}&Uniform(1 - 11) \\
    \enddata
    \caption{Parameters used for GPI \kband \prt  free retrieval}
    \label{tab:prt_params}
\end{deluxetable*}

    

\par The main difference between the retrieval results and forward model results is the surface gravity. The GPI \kband free retrieval favored a value of 3.57 $\pm$ 0.74 cm s$^{-2}$ dex, marginally inconsistent to 1$\sigma$ with the preferred value of the forward model of 4.55\plusminus{0.17}{0.22} cm s$^{-2}$ dex. We also see the favored planetary radius is smaller than expected and up against the prior space boundary. We ran multiple retrievals with a varied prior range and found that the retrieval always favored the smallest possible planet radius, even in a case where the radius was not physically-motivated, i.e. 0.7 \Rj. The retrieved temperature is consistent with the forward model results. 
In addition, we also see a mean retrieved Fe grain size of $28 \pm 14 \mu$m and MgSiO$_3$ mean grain size of $48\pm24 \mu$m, which is larger than the \texttt{PHOENIX} characteristic grain size of 1 $\mu$m. The retrieved C/O ratio is 0.40$\pm$0.19, which is marginally consistent to 1$\sigma$ with our custom C/O model grid at 0.57\plusminus{0.02}{0.02}, and much less constrained. We see a large inconsistency between the retrieved surface gravity on low-resolution GPI data and the analysis of the high-resolution KPIC data with a self-consistent atmospheric grid. The low retrieved surface gravity, \logg $= 3.57 \pm 0.74$ is not only not as constrained, but also comes from a retrieval with a corresponding unphysically small companion radius, likely stemming from a mismatch in the cloud model which is common among young, dusty companions \citep{Marois_2008, bowler_hr8799_2010_osiris, Barman_2011, 2011ApJ...729..128C, Marley_Saumon_Cushing_Ackerman_Fortney_Freedman_2012, Hoch_2022}. Using the retrieved effective temperature and surface gravity, we compute the Bayes' factor, $B$ (Equation \ref{eq:bayesfactor} and surrounding discussion), of these parameters against our most probable values from the MCMC analysis using the custom \texttt{PHOENIX} model grid in Section \ref{sec:tlimit}. For these models, $B=0.00028$, strongly rejecting the \logg favored by the retrieval \citep{Jeffreys_1983}. Varying \logg changes the absorption line shape as well as the CO-bandhead shape, which our high-resolution data is highly sensitive to as KPIC resolves the individual absorption features. The small companion radius produced by the retrieval coupled with the sensitivity of the high-resolution data to changes in \logg leading to rejection of the lower \logg from the retrieval motivates an exploration of retrievals with more confined prior spaces in accordance with the most probable parameter ranges from our analysis with the self-consistent \texttt{PHOENIX} atmosphere model grid.

\par
\begin{figure*}
    \centering
    \includegraphics[width=\linewidth]{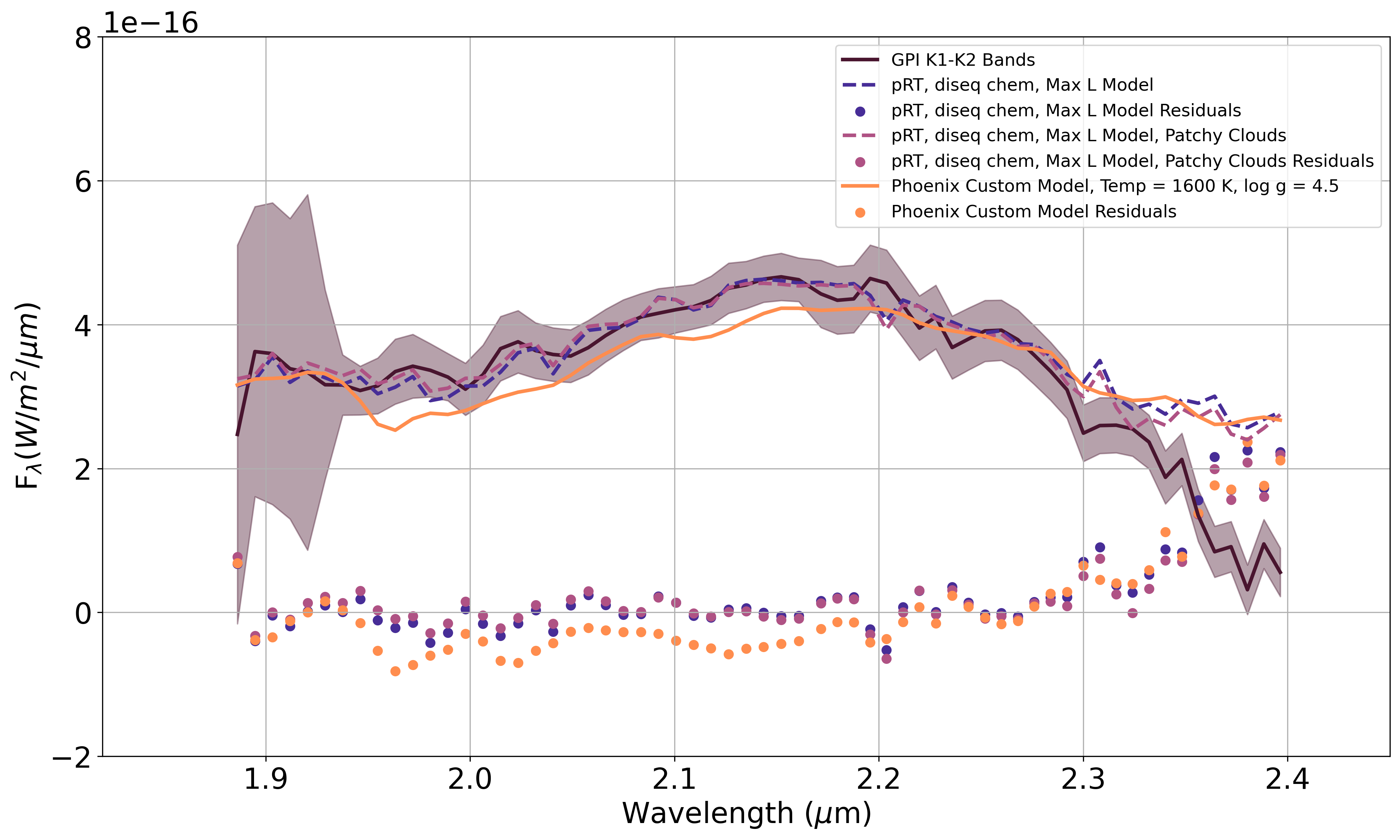}
    \caption{Gemini Planet Imager \kband data with the spectra generated by the maximum likelihood values from petitRADTRANS disequilibrium chemistry free retrievals and the best fit custom Phoenix models. Model parameters are listed in Table \ref{tab:prt_limited_retrieval}.}
    \label{fig:prt_bestfits}
\end{figure*}

 We also ran two retrievals on GPI \kband data alone with a limited prior space reflective of our findings from the high-resolution KPIC data. We limit the temperature and \logg prior range to $\pm 2 \sigma$ of the posterior distribution from Section \ref{sec:tlimit}. For these retrievals, we use two of the built-in emission retrieval models, \texttt{emission\_model\_diseq} and \texttt{emission\_model\_diseq\_patchy\_clouds}, the latter of which allows for the parameterization of patchy clouds where the parameter ``patchiness'' indicates the fraction of cloud coverage, 0 being clear and 1 being completely cloudy. The results from these retrievals and the prior probability spaces are shown in Table \ref{tab:prt_limited_retrieval} and the spectra generated by the maximum likelihood values of each retrieval are shown in Figure \ref{fig:prt_bestfits}. 

\begin{deluxetable*}{cccc}
    \tablehead{\colhead{Parameter}&\colhead{Dis-equilibrium Chemistry Model}&\colhead{Dis-equilibrium Chemistry Model, Patchy Clouds}&\colhead{Prior Range}}
    \startdata
        \logg [cm s$^{-2}$ dex] &  4.51$\pm$0.23 & 4.50$\pm$0.23 & Uniform(4.11 - 4.89)\\  \hline
        Radius [R$_{Jup}$] &  1.20 $\pm$0.20 & 1.21 $\pm$0.22 & Uniform(0.9 - 2.2)  \\  \hline
        Effective Temperature [K] & 1649$\pm$41 & 1638$\pm$39 & Uniform(1570 - 1712)  \\   \hline
        $\alpha$ & 1.32$\pm$0.25 & 1.41$\pm$0.27 & Uniform(1 - 2)  \\ \hline
        log($P_{quench}$) [bar] & -1.3$\pm$2.9 & -1.4$\pm$2.8 & Uniform(-6 - 3)  \\ \hline 
        log($\Delta$) & 0.716$\pm$0.061 & 0.708$\pm$0.066 & Uniform(0 - 1.0)  \\  \hline
        Fe/H [dex] & -0.10$\pm$0.60 & -0.09$\pm$0.54 & Uniform(-1.5 - 1.5)  \\ \hline
        C/O & 0.40$\pm$0.19 & 0.43$\pm$0.17 & Uniform(0.1 - 1.6)  \\  \hline
        $\sigma_{log\_norm} [\mu$m] & 1.97$\pm$0.58 & 2.00$\pm$0.56 & Uniform(1 - 3)\\ \hline
        log($k_{zz}$) [cm$^2$ s$^{-1}$] & 10.3$\pm$2.2 & 9.2$\pm$2.3 & Uniform(5 - 13)  \\ \hline
        f$_{sed}$ & 7.0$\pm$2.6 & 6.9$\pm$2.7 & Uniform(1 - 11)  \\   \hline
        patchiness & --- & 0.44$\pm$0.26 & Uniform(0-1) \\
    \enddata
    \caption{Parameters used for GPI \kband \prt  limited effective temperature and \logg  retrieval}
   \label{tab:prt_limited_retrieval}
\end{deluxetable*}

\par As expected, the most notable changes to the posterior distributions of the retrieval are the surface gravity and radius of the planet. The retrieved values of $4.51\pm0.23$ and $4.50\pm0.23$ (patchy cloud model) are consistent with the forward model's favored value of 4.55\plusminus{0.17}{0.22}. We also see a much more physically-motivated radius of $1.20\pm0.20$\,\
\Rj and $1.21\pm0.22$\,\Rj  (patchy cloud model). All other parameters are consistent to the less constrained retrieval, though notably the unitless pressure power law scaling parameter, $\Delta$, is much better confined at a value of $0.716\pm0.061$ and $0.708\pm0.066$ (patchy cloud model) compared to $0.56\pm0.22$ in the first retrieval. 

\par
\begin{figure*}
    \centering
    \includegraphics[width=\linewidth]{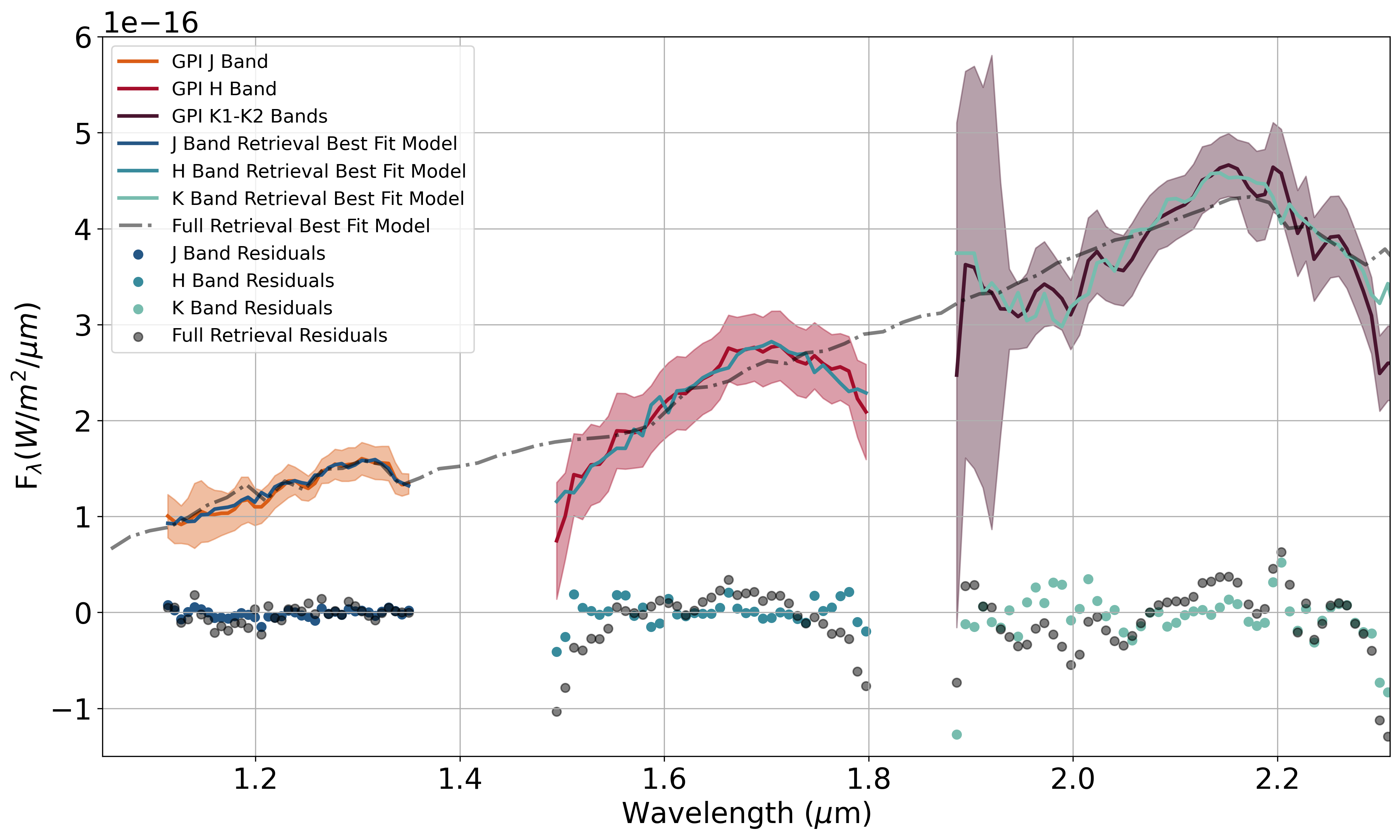}
    \caption{We plot the GPI \jband, \hband, and \kband spectra from \citealt{Ward-Duong_GPI_spec} along with the best-fit retrieval model from \prt including all 3 spectral bands. In addition, we show the best-fit models from GPI retrievals using only one spectral band at a time while using the same PT parameterization and prior range.}
    \label{fig:retrievalresults}
\end{figure*}

 Although the GPI \kband \prt retrieval results are broadly similar to our forward model analysis of high resolution KPIC \kband  data in Section \ref{sec:tlimit}, the retrieved model still does not match well to \jband and \hband. We use the same \prt framework to do retrievals on both \jband and \hband and find strong disagreement between the bands, especially in \hband. When fitting all three wavelength bands together, the free retrieval is able to broadly fit the overall spectral energy distribution, but does not reproduce the interband features present in the observed spectrum. The full retrieved spectrum, as well as the best fit \kband, \jband, and \hband individual best fit free retrieval models are shown in Figure \ref{fig:retrievalresults}.

\section{Bulk Properties from Evolutionary Models}

\par
\begin{figure*}
    \centering
    \subfigure[Age and Mass]{\includegraphics[width=0.45\linewidth]{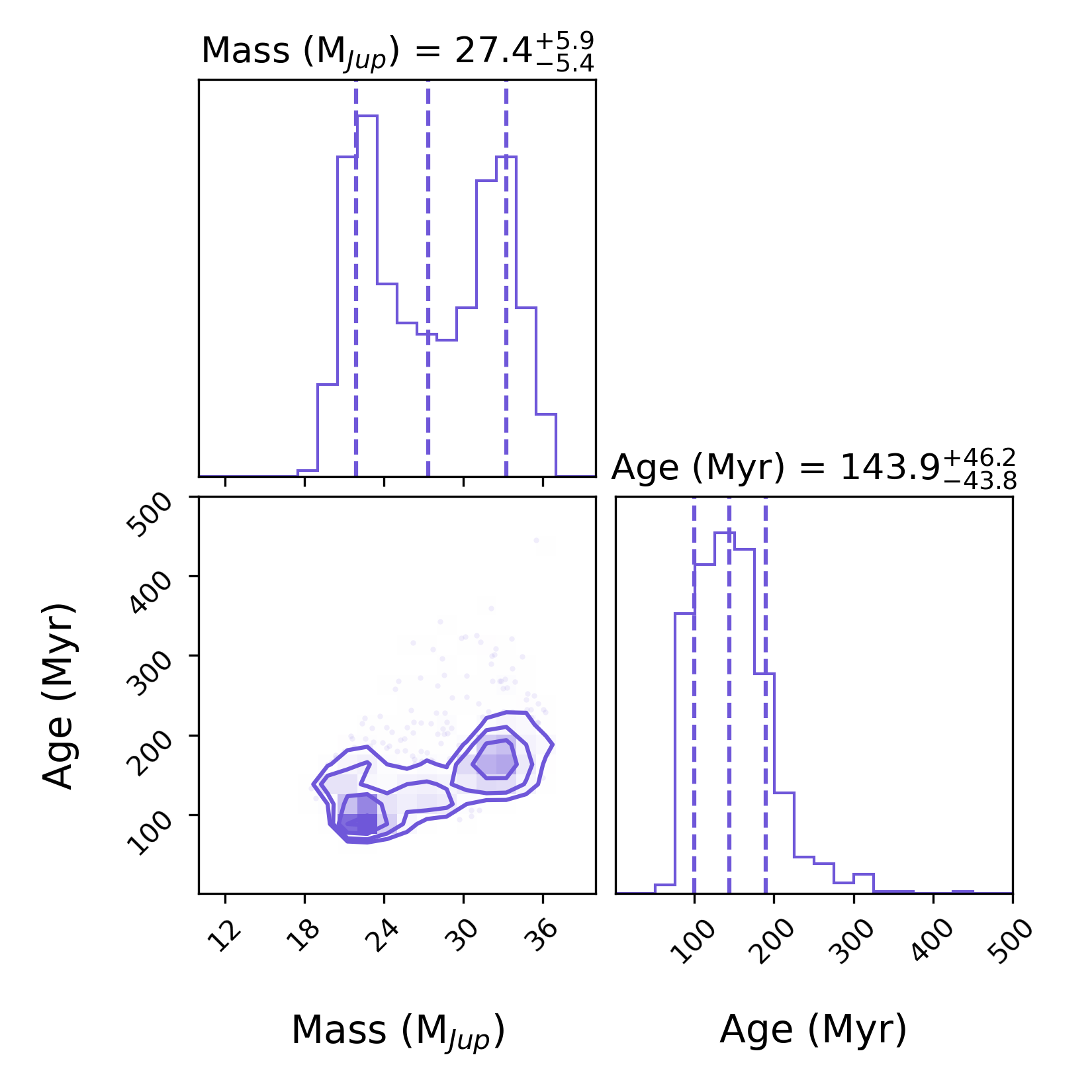}}
    \qquad
    \subfigure[Mass and Radius]{\includegraphics[width=0.45\linewidth]{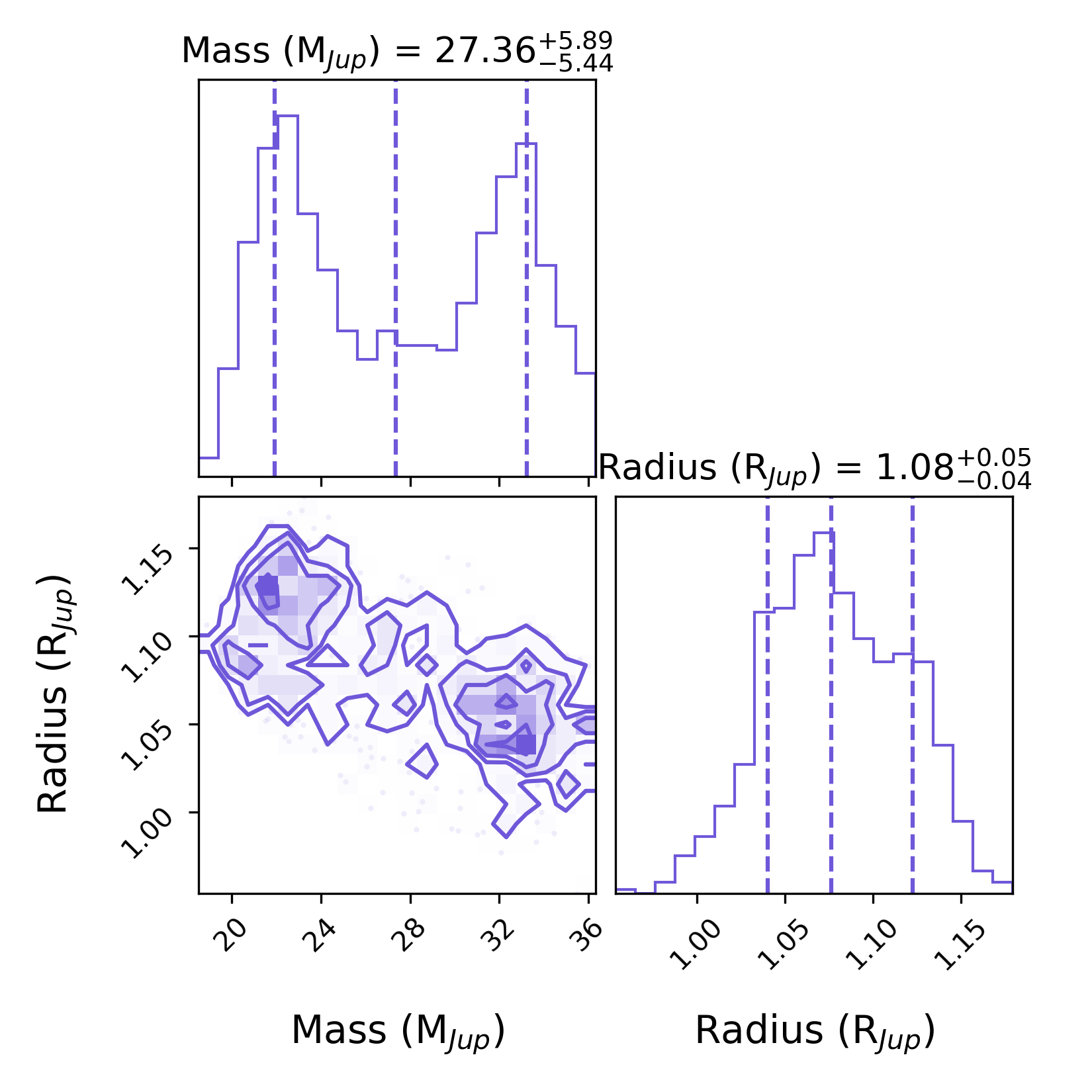}}
    \caption{(a): Interpolated age and mass distributions from CBPD23 evolutionary tracks using the full posterior distribution from forward modeling effective temperature and surface gravity with \texttt{PHOENIX} atmospheric models. The bimodal mass distribution corresponds to a bimodal temperature distribution in the forward model posterior. (b): The same posterior corresponds to a radius of 1.08 \plusminus{0.05}{0.04} \Rj.}
    \label{fig:evolution_full}
\end{figure*}

 Given the posterior distribution of the temperature and gravity parameters from our modeling, we can use evolutionary models to estimate the mass and age of \planet. We choose evolutionary tracks from \citealt{chabrier_evo_models_2022}, hereafter CBPD23, which span from 1.048 \Mj to 78.6 \Mj and 1 Myr to 10 Gyr in age. We interpolate these models onto a linear temperature and gravity grid and using the posterior distribution from our MCMC with the custom \texttt{PHOENIX} atmospheric model grid (see Section \ref{sec:mcmcresults}), we select pairs of effective temperatures and \logg s and plot the corresponding age-mass values and radius-mass values in Figure \ref{fig:evolution_full}.  We find a companion mass from evolutionary models of 27.4\plusminus{5.9}{5.4} \Mj in a bimodal distribution. We also use CBPD23 to calculate the expected radius of the companion: 1.08 \plusminus{0.05}{0.04} \Rj when considering the full temperature grid in Figure \ref{fig:evolution_full}. This companion mass is largely consistent with the literature, and notably consistent with the dynamical mass of \planet \xspace reported by \citealt{Hinkley_2023_c} of 26.2\plusminus{3.7}{3.6}.

\par
\begin{figure*}
    \centering
    \subfigure[Age and Mass (Temp Limit)]{\includegraphics[width=7cm]{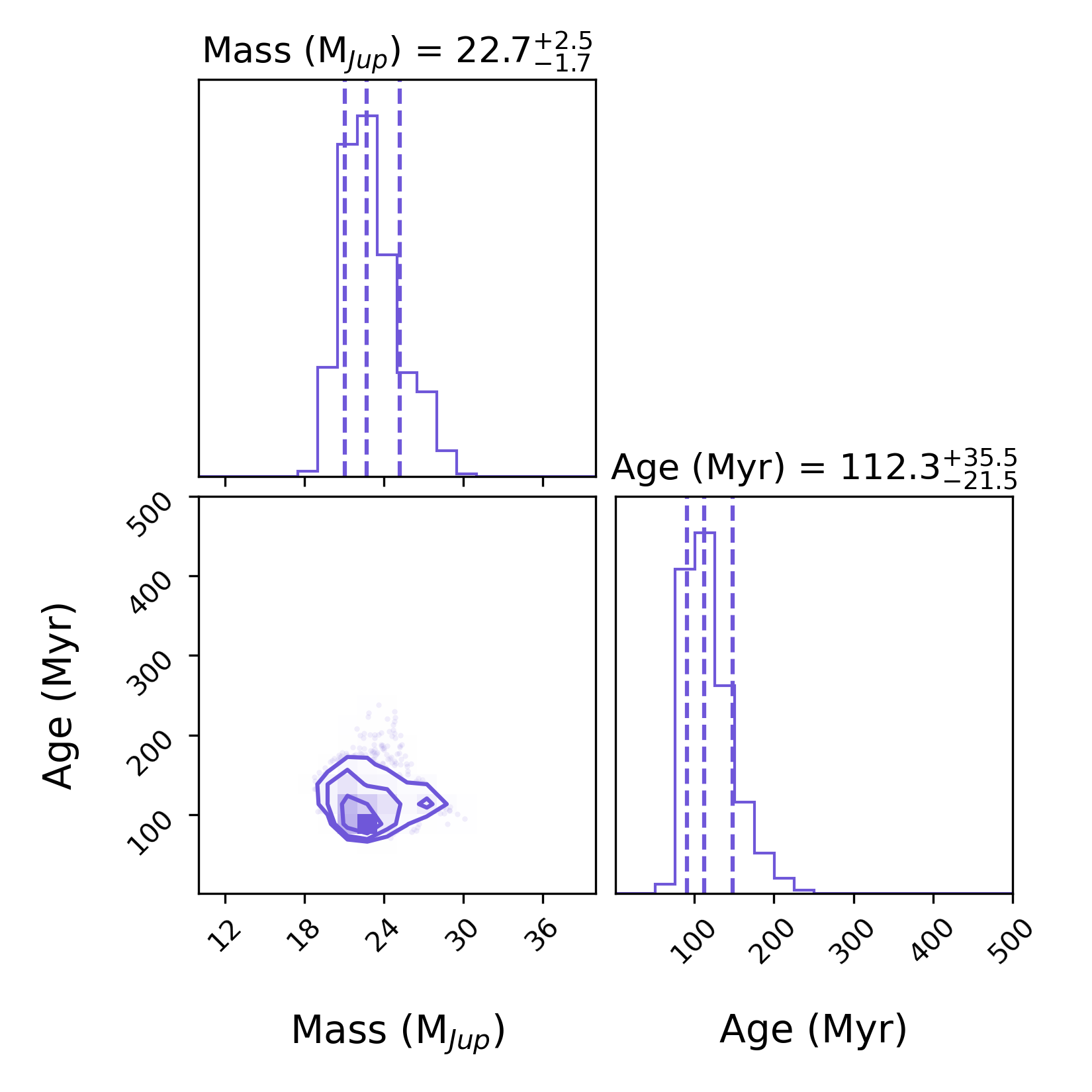}}
    \qquad
    \subfigure[Mass and Radius (Temp Limit)]{\includegraphics[width=7cm]{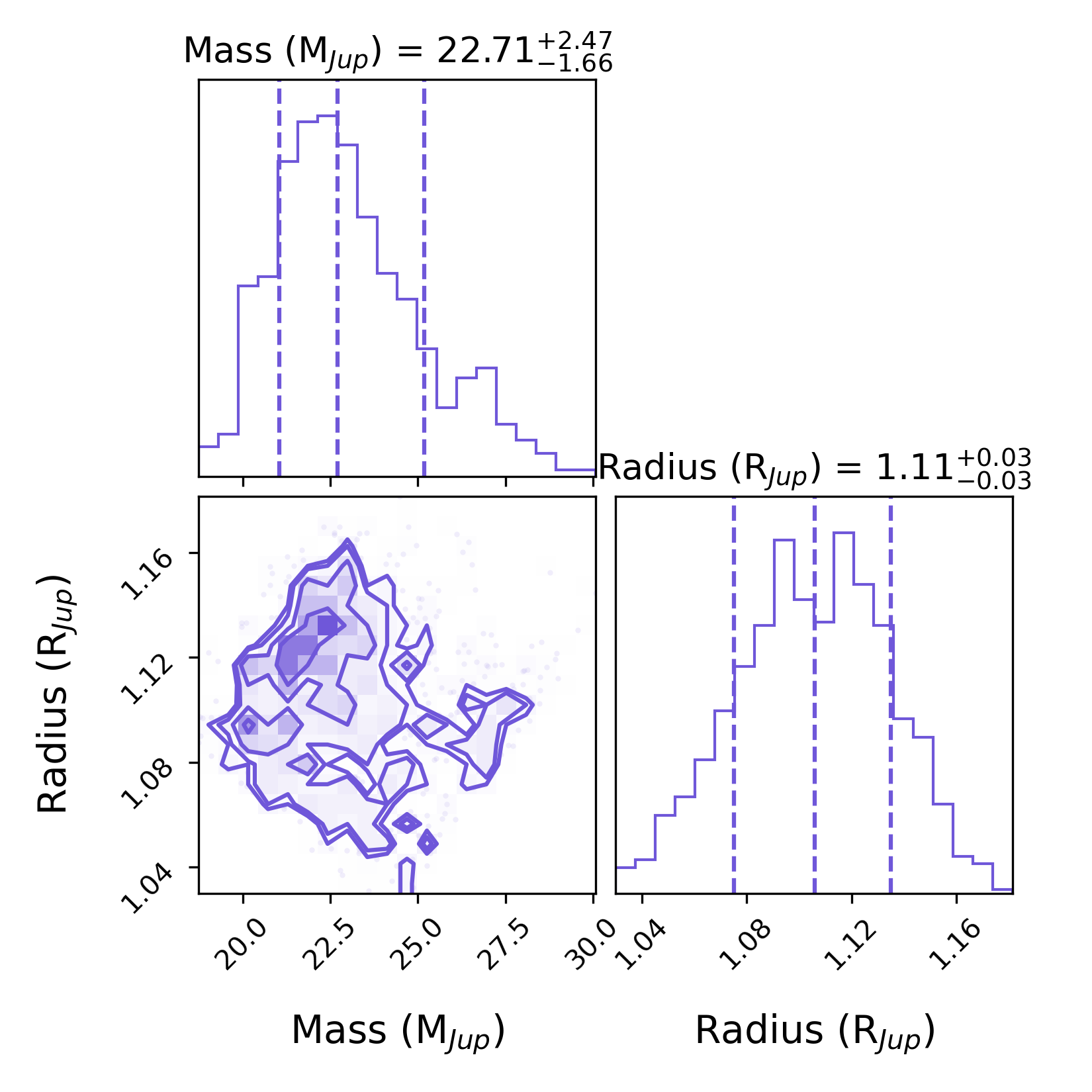}}
    \caption{(a):Interpolated age and mass distributions while limiting the temperature prior space to below 1750K. The bimodality of the mass distribution disappears in this temperature range while the age of the object remains constant with the full model grid. (b): The bimodal mass distribution shifts to the lower mass value, while the radius increases slightly, though still consistent with the full grid.}%
    \label{fig:evolution_crop_t}%
\end{figure*}

 However, when we limit our temperature range to T$_{eff} < 1750$ K and below, motivated by our low-resolution analysis of archival GPI data, the resulting age and mass distribution completely eliminates the posterior peak of higher mass. The mass from evolutionary models drops to 22.7\plusminus{2.5}{1.7} \Mj. This mass is still consistent to $1\sigma$ with the dynamical mass reported by \citealt{Hinkley_2023_c}. The lower mass solutions also correspond to younger, larger radius solutions. We extend the same treatment to the radius of the companion from evolutionary models, with the results of both posterior distributions shown in Figure \ref{fig:evolution_crop_t}. We see the most likely radius move from 1.08\plusminus{0.05}{0.04} \Rj to 1.11\plusminus{0.03}{0.03} \Rj and the age of the companion decrease from 144 Myr to 112\plusminus{36}{24} Myr, though both the radius and age are consistent to $1\sigma$. 

\section{Orbital Analysis} \label{orbital_analysis}

\par Fitting our high-resolution spectrum allows us to add two radial velocity points to the HD 206893 B astrometry measurements. We use the orbit fitting code \texttt{orbitize!} \citep{Blunt_2020_orbitize} to perform a joint orbit fit analysis of the HD 206893 system. We include the published relative astrometry, \textit{Gaia} astrometry, and stellar RV data \citep{Milli_discovery,Hinkley_2023_c} with the addition of our 2020 and 2022 RV values. We use the MCMC sampler with 1,000 walkers, 20 temperatures, 10,000 burn-in steps and a thin factor of 10. Following the mathematical formulation outlined in \citealt{Lacour_2021}, we also consider the secular perturbations of companion c on companion B, as considering the effect of this perturbation has improved residuals in Gravity-precision data \cite{Lacour_2021}. Our final orbit fit is shown in Table \ref{tbl:jointfitposteriors}. 
We note that our fit is consistent with the one reported in \citealt{Hinkley_2023_c}, but presents a higher eccentricity for HD 206893 b of 0.27\plusminus{0.04}{0.23} (they report an eccentricity of $0.14\pm 0.05$). However, for both HD 206893 B and c, the eccentricity solutions allow for near-circular orbits.

\begin{deluxetable}{lccc}
\tablecaption{Joint Fit for HD 206893 B and c} \label{tbl:jointfitposteriors}
\tablewidth{20pt}
\tablecolumns{4}
\tabletypesize{\scriptsize}
\tablehead{
    \colhead{Parameter} & \colhead{B} & \colhead{c}
}
\startdata
$a$ (au) & 8.93\plusminus{1.41}{0.19} & 3.62\plusminus{0.35}{0.42} \\
$e$ & 0.27\plusminus{0.04}{0.23} & 0.38\plusminus{0.09}{0.10} \\
$i$ ($^\circ$) & 155.96\plusminus{2.42}{17.76} & 148.91\plusminus{8.92}{5.26} \\
$\omega$ ($^\circ$) &203.59\plusminus{13.55}{161.75} & 27.1\plusminus{7.4}{8.6} \\
$\Omega$ ($^\circ$) & 104.34\plusminus{143.37}{13.15} & 196.61\plusminus{93.39}{189.51} \\
$\tau$ & 0.31\plusminus{0.16}{0.01} & 0.65\plusminus{0.20}{0.04} \\
$M$ ($M_{\rm Jup}$) & 23.88\plusminus{0.86}{2.47}& 10.31\plusminus{2.39}{2.22} \\
\hline
\multicolumn{3}{c}{\textbf{Shared Parameters}} \\
\hline
Parallax (mas) & \multicolumn{2}{c}{24.45\plusminus{0.22}{0.08}} \\
$M_*$ ($M_\odot$) & \multicolumn{2}{c}{1.32\plusminus{0.13}{0.04}} \\
\enddata
\end{deluxetable}

\subsection{Stability}

\begin{figure*}
    \centering
    \includegraphics[width=\linewidth]{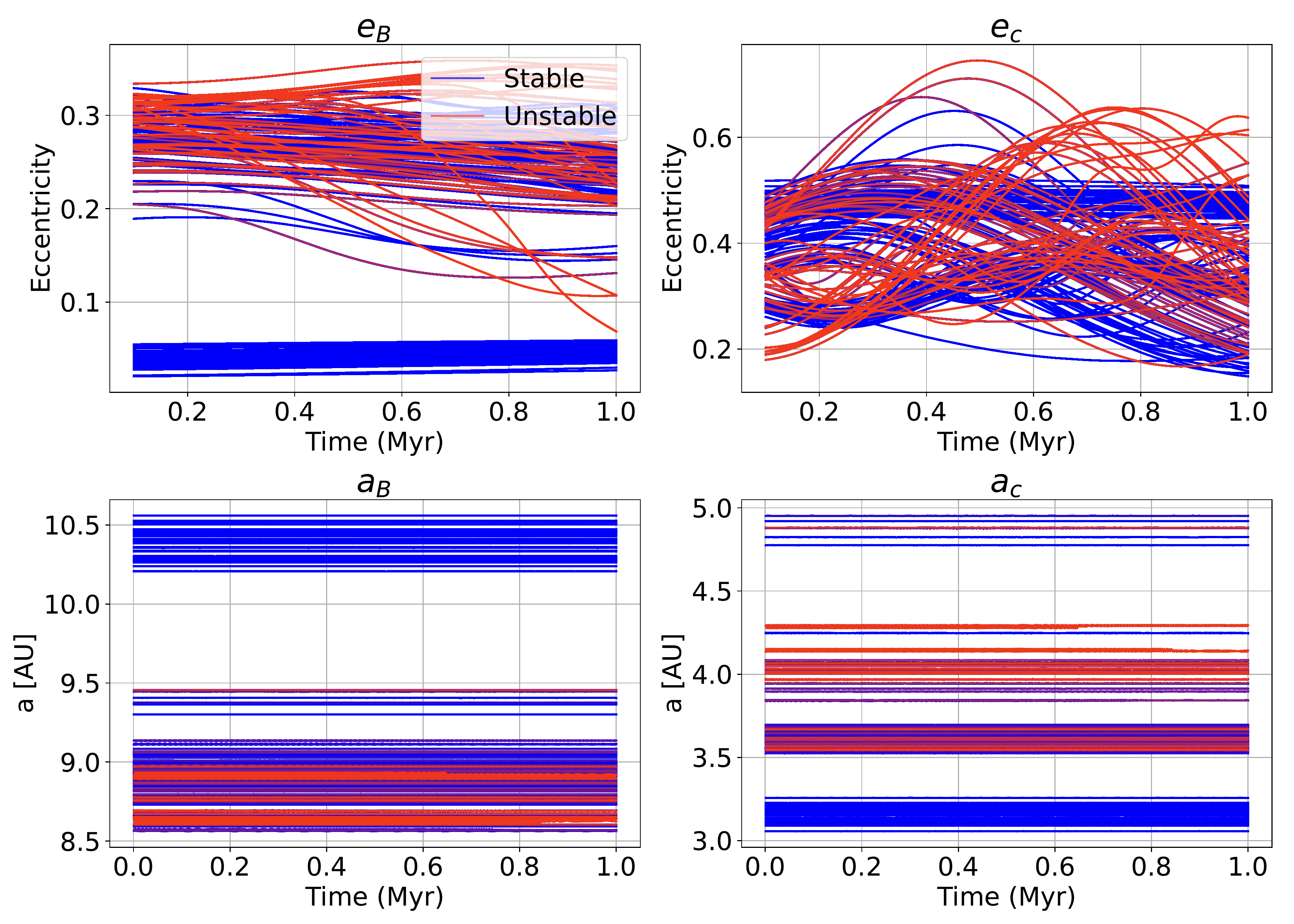}
    \caption{Semi-major axis and eccentricity over time for random draws for HD 206893 B and c for 1 Myr integration. We find that there is no strong preference for circular/elliptical orbits or specific semi-major axis values. Configurations are stable in about $\sim$ 95.8\% of cases.}
    \label{fig:1myr}
\end{figure*}

\begin{figure*}
    \centering
    \includegraphics[width=\linewidth]{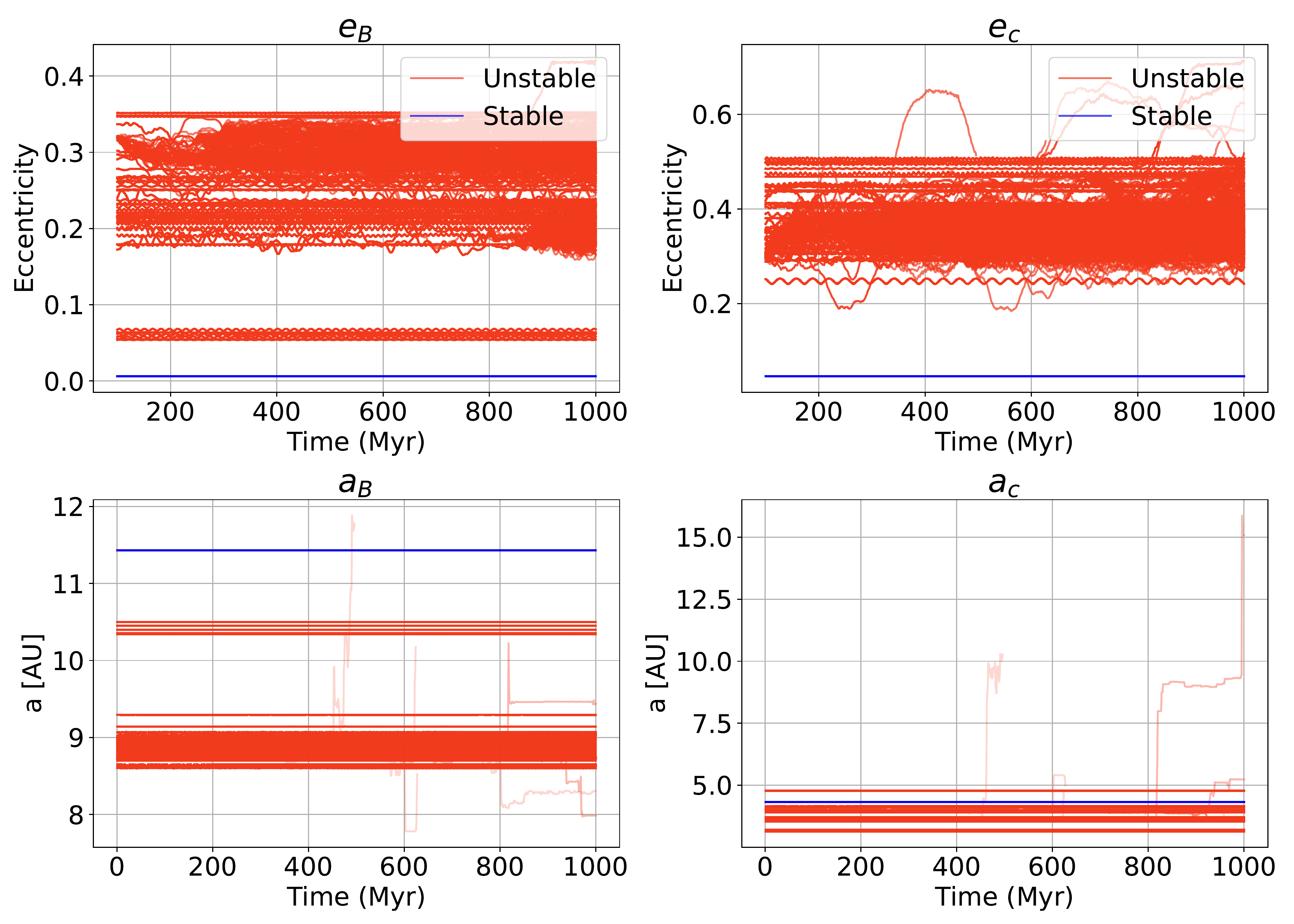}
    \caption{Semi-major axis and eccentricity over time for HD 206893 B and c for 1 Gyr integration. We find that there is a strong preference for circular orbits and for a larger semi-major axis for B and c. Configurations are stable in $\sim$ 0.8\% of cases.}
    \label{fig:1gyr}
\end{figure*}

After deriving updated orbital parameters with the new RV points, we investigate the stability of the system. We use the N-body code \texttt{REBOUND}'s WHFast integrator \citep{Rein_Tamayo_2015} with a timestep of 0.1 years. We obtain 10,000 random draws from the full orbital posteriors, which include B and c's orbital parameters along with all component masses. For each draw, we track the semi-major axis and eccentricity over time for both objects, along with the Mean Exponential Growth of Nearby Orbits (MEGNO) value \citep{Maffione_Giordano_Cincotta_2011} over time. Stable configurations are expected to obtain final MEGNO values between 1.95 and 2.05 (e.g., \citealt{Gozdziewski_2014}). We do this for two integration times: 1 Myr and 1 Gyr. We find that the system is stable in 95.8\% of draws for 1 Myr integration time and in 0.8\% of draws for 1 Gyr integration time. The configurations that remain stable for 1 Gyr strongly favor low eccentricities (e$<$0.1) for both objects, although these solutions are disfavored by the joint orbit fit (see Figures \ref{fig:1myr} and \ref{fig:1gyr}). The stable configurations for 1 Gyr all have the same same semi-major axis values, of $a_B = 11.43$ and $a_c = 4.32$ AU. The eccentricities are 0.005 ($e_B$) and 0.038 ($e_c$).

Given the strong preference for circular orbits in the long-term stability analysis (and the disfavoring of these solutions in the joint orbit fit) we also assess individual orbit fits for both objects using observable-based priors \citep{O'neil_obs_based_priors_2019}, which aim to mitigate high-eccentricity biases in orbit fits where there is a short orbital arc. We present our results in Appendix \ref{orbit_appendix}.

\section{Discussion}\label{sec:discussion}

\par The bulk atmospheric parameters reported in Section \ref{sec:mcmcresults} are the first produced from a high-resolution spectroscopic analysis of \planet. Our results from the high-resolution data differ somewhat from previous works. Notably, the data prefer modestly higher temperatures than reported by \citealt{Kammerer_Gravity_K_band} and \citealt{Hinkley_2023_c}, who measured 1589\plusminus{13}{22} K and 1429.2\plusminus{5.6}{6.2} K respectively. While we are near agreement with values reported by \citealt{Kammerer_Gravity_K_band} using a dusty BT-Settl model set, to 1$\sigma$, our corresponding surface gravity measurement is 0.76 dex higher. On the other hand, our maximum likelihood value of \logg is consistent with \citealt{Hinkley_2023_c}  while we see an effective temperature 300 K hotter.

\par Our high-resolution spectra also allow us to investigate the companion's rotational velocity for the first time. We found a projected spin velocity of \vsini = 9.8 \plusminus{1.9}{2.0} km s$^{-1}$. To quantify the significance of our spin measurement posterior distribution (compared to a spin non-detection where the instrumental broadening inherent to Keck/NIRSPEC masks the true spin velocity), we compute the Bayes' factor, $B$, which we use to compare the relative probability of two models as in \citealt{Wang_HR8799_2021} (see Equation 9 within and subsequent discussion). 
\begin{equation}
    B \equiv \frac{P(D|M_1)}{P(D|M_2)} = \frac{P(M_1|D)P(M_2)}{P(M_2|D)P(M_1)}
    \label{eq:bayesfactor}
\end{equation}
A Bayes' factor of $B<0.05$ shows ``very strong'' evidence of the second model being favored over the first, and the first model can only be rejected under the condition of $B<0.01$ \citep{Jeffreys_1983}. Since we assume an equal likelihood of both models given the data, $B$ is simply the relative probability between two models. For the case of no spin, we find $B = 0.0122$ for the 2020 epoch, and cannot fully reject the no-spin model (equivalent to not detecting evidence of spin broadening). In the 2022 epoch, $B = 2.61$ and the no-spin model is favored over the spin-broadened model. The posterior distribution of \vsini is inconsistent with a spin of 0 km s$^{-1}$ by 4.9 $\sigma$, but $B$ gives a more conservative estimate of our ability to measure \vsini with a dataset of this SNR, and a value of projected spin this close to the instrumental broadening resulting from a resolving power of 35,000, estimated to be $\sim5-7$ km s$^{-1}$. From the aggregate dataset, the data is inconclusive regarding the companion's true projected spin. However, from the MCMC results, we establish an upper limit on \vsini of 15.5 km s$^{-1}$, $3-\sigma$ from the mean of the projected spin distribution.
We found the break-up velocity of the companion to be $214 \pm 27$ km s$^{-1}$, derived from the mass determined from planetary evolution models. Therefore, the upper limit of the projected rotation of the companion is $0.072 \pm 0.024 $ $v_{break}$. This rotation rate is in alignment with the trend reported in \citealt{Wang_HR8799_2021} where higher-mass companions have a slower rotation rate relative to their break-up velocity compared to lower-mass companions, pointing to more efficient braking in higher-mass objects. 

\par The determination of bulk atmospheric properties of effective temperature and \logg allow for the determination of a planetary age and mass from evolutionary models based on our high-resolution data. Planetary ages and masses allow for further comparison to contemporary measurements using different techniques. \citealt{Dupuy_Liu_2017_dynamicalmasses} finds good agreement between between atmospheric evolution models of \cite{BHAC2015} (BHAC15 hereafter) and the measured dynamical masses of brown dwarfs with M $\geq$ 60 \Mj. Extending the use of these evolutionary models to lower mass, lower temperature objects, we turn towards CBPD23 evolutionary models as they are a continuation of the BHAC15 model set and extend to lower effective temperatures, and with an updated equation of state that better models the interaction between H and He in giant planets. 

\par Specifically, our derived mass of 22.7 \plusminus{2.5}{1.7} \Mj is consistent with the dynamical mass of \planet \xspace reported by \citealt{Hinkley_2023_c}  of 26.2 \plusminus{3.7}{3.6} \Mj and with our dynamical mass fit of 23.88 \plusminus{0.86}{2.47} \Mj. Our derived age of 112 \plusminus{36}{22} Myr is consistent with estimates from \citealt{Hinkley_2023_c} as well as ranges provided by \citealt{Kammerer_Gravity_K_band} based on evolutionary tracks from AMES-Cond \citep{Baraffe_amescond_2003} and their results from using a dusty \texttt{BT-Settl} modelset in a similar analysis (see \citealt{Kammerer_Gravity_K_band} for complete grid details), in the same family of models as CBPD23. We see a growth in planetary radius with a decrease in temperature, meshing well with the contention that \planet \xspace has a puffy, dusty atmosphere causing reddening in its spectrum through dust extinction in \jband and \hband.

\par Our analysis also includes a free retrieval of GPI low-resolution spectra of HD 206893 B. While the model fit to \kband data alone generally fits well, running a free retrieval on \jband, \hband, and \kband simultaneously does not produce consistent results, with the full retrieval, and individual bands shown in Figure \ref{fig:retrievalresults}. These fits are still better than the fully consistent models in \jband and \hband, which still show an order of magnitude of extinction in \jband and slightly less extinction in \hband while fitting well in \kband. Most notably we still see a retrieved radius of $2.16 \pm 0.33$ \Rj and subsolar metallicity of $-1.391\pm 0.066$. Both retrieved parameters show large differences from the retrieved \kband data alone, as well as the MCMC results from our \texttt{PHOENIX} atmospheric model grid, and notably a metallicity that is effectively impossible giving the system's youth. 


\par Both our full spectrum retrieval and the atmospheric models appear to fail to describe the cloud and dust grain parameters necessary to achieve the observed level of extinction. All retrievals point to a dust grain distribution with characteristic size of $\sim 30 \mu$m and a lower temperature/surface gravity in tandem with a larger radius to fit the observed reddening. Recent work has shown that across the L/T transition, higher temperature, higher \logg atmospheres with smaller dust grains and higher clouds can produce similar spectra as objects move from early L to late T-type objects \citep{Brock_2021}, and though these parameters are not preferred by the retrievals on low resolution data, we see good agreement between the overall morphology of the K-band spectra and the higher temperature/\logg models. 

\begin{deluxetable*}{cccccccc}
    \tablehead{\colhead{Temperature (K)}&\colhead{\logg}&\colhead{Radial Velocity (km s$^{-1}$)}&\colhead{Projected Spin (km s$^{-1}$)}&\colhead{Mass (\Mj)}&\colhead{Age (Myr)}&\colhead{C/O ratio}&\colhead{Ref.}}
    \startdata
        1200-1380&  ---&  ---&  ---& 24-73  &  ---&  ---& 1\\ \hline
        1300-1600 &3.5-5.0 & ---& ---&15-30  &50-700 & ---&2\\\hline
        ---& ---& ---& ---& 10 \plusminus{5}{4}& ---& ---&3\\ \hline
        1200-1800 &  3.0-5.0&  ---&  ---&   ---&$\sim$50 (kinematics)&---& 4 \\\hline
        ---& ---& ---& ---& 12 - 78& $< 100$& ---&5\\ \hline
        1589\plusminus{13}{22}&3.83\plusminus{0.38}{0.14}&  ---&---& ~5 - 30&3-300& ---& 6*\\\hline
        1216\plusminus{13}{17}& 2.87\plusminus{0.63}{0.47}& ---& ---& 0.83\plusminus{2.71}{0.55}& ---& 0.82\plusminus{0.04}{0.19}&6$^{\dagger}$\\ \hline
        1429.2\plusminus{5.6}{6.2} &  4.66\plusminus{0.04}{0.04} &  ---&  ---& 26.2 \plusminus{3.7}{3.6}\vspace{.15cm} &$\sim 150$&---& 7\\ \hline
        \hline
        1774\plusminus{136}{159} \vspace{.2cm}& 4.70\plusminus{0.17}{0.27}& -9.10\plusminus{0.81}{0.77}& 15.6$^{\P}$ (9.3\plusminus{2.1}{2.3}) & 27.4\plusminus{5.9}{5.4} & 144\plusminus{47}{44} & 0.57\plusminus{0.02}{0.02} &This work\\ \hline
        1634 \plusminus{72}{38}&4.55\plusminus{0.17}{0.22}&-9.00\plusminus{0.79}{0.77}& 15.5$^{\P}$ (9.8\plusminus{1.9}{2.0})  & 22.7\plusminus{2.5}{1.7} & 112\plusminus{36}{22}& 0.57\plusminus{0.02}{0.02} &This work $^{\ddag}$ \\
    \enddata
    \caption{Amalgamated Bulk Physical Parameters of HD 206893 B from the literature and this work. 1: \citealt{Milli_discovery}; 2:\citealt{Delorme_2017}, 3:\citealt{Grandjean_2019}; 4: \citealt{Ward-Duong_GPI_spec}; 
        5:\citealt{Meshkat_near_to_thermal};
        6*: \citealt{Kammerer_Gravity_K_band}, Best-Fit BT-Settl Dusty Model with Stellar RV included in orbit fit; 6$^{\dagger}$: \citealt{Kammerer_Gravity_K_band}, \texttt{petitRADTRANS} Free Retrieval; 7:\citealt{Hinkley_2023_c} ;$^{\ddag}$Maximum temperature limited to 1750K to motivate a reasonable companion radius; $^{\P}$ $3-\sigma$ upper limit on \vsini with the statistical measurement and uncertainty reported for completeness.}
   \label{tab:all_works}
\end{deluxetable*}

\par  As noted in Table \ref{tab:all_works}, we report the C/O of \planet \xspace from a self-consistent atmospheric model grid at $0.57 \pm 0.02$. Recent work reported the C/O ratio of HD 206893 A to be super-solar ($0.81 \pm 0.14$) to $2\sigma$ when fitting high-resolution spectra with \texttt{PHOENIX-C/O} models due to an enhanced carbon abundance, though using the equivalent widths method on the same spectrum yields a C/O value consistent with both our measurement and the solar value ($0.69 \pm 0.35$)\citep{baburaj_2024_stellar_metallicities}. 
\citealt{Kammerer_Gravity_K_band} reports a C/O ratio of  0.65 (plain) to 0.75 (dusty) using the self-consistent Exo-REM model grid \citep{Baudino_2015_exorem,Charanay_2018_exorem}, and a value of 0.82 \plusminus{0.04}{0.19} from a \prt free retrieval, which is consistent compared to \citet{baburaj_2024_stellar_metallicities}, but comes from a free retrieval with a companion mass of 0.83 \plusminus{2.71}{0.55} \Mj, which is inconsistent with measured dynamical mass. \citealt{Hoch_2023} found that gas giants with M $>$ 4\Mj have an approximately solar C/O ratio, independent of their separation, a finding that is consistent with our observation. We aim to use C/O ratio as an indicator of formation location and as evidence towards a planetary formation pathway, to differentiate between a formation via pebble accretion vs disk fragmentation. We cannot determine the formation pathway of this system from only the C/O ratio alone for several reasons. First, the chemical evolution of the protoplanetary disk is dynamic and changes the C/O ratio of material accreted by a companion if the companion is undergoing migration \citep{2020A&A...640A.131M}. Second, it has recently been shown that the measured C/O ratio cannot help to distinguish an in-situ massive planet formation from a formation and subsequent migration, though a giant planet that has been enriched through accretion of solid material from the protoplanetary disk is expected to have a slightly sub-stellar C/O ratio \citep{Turrini_2021_giantplanetCO}. The metallicity retrieved from GPI data, consistent with solar, does not rule out a core accretion scenario, as the gas accreted after a core forms could be depleted in metals due to planetisimal formation at that location in the disk \citep{Helled_2014prpl.conf..643H}. 
\par The presence of a debris disk exterior to \planet \xspace plus the existence of an interior companion creates an interesting possibility of a planet-like formation via core accretion. If \planet \xspace had formed from disk instability at larger separation( $\sim100$ au) and migrated inward, it is difficult to explain the current presence and morphology of the intact debris disk around HD 206893 with its inner edge at 30 au. Indeed, if \planet \xspace formed much farther out and migrated inward, one would expect evidence of metal enhancement from the planetisimal bodies in the debris disk as well. Massive planet formation via core accretion is sensitive to the metallicity of the planet-forming disk and the location of formation, with the area of highest formation efficiency around 5-10 au around more massive stars \citep{Helled_2014prpl.conf..643H}. Though we do not retrieve high metallicity in the atmosphere of \planet, this could be due to the amount of metal-depleted gas accreted by the planet core after its formation, resulting in a companion showing no enhancement in metallicity and stellar C/O ratio from an efficient planetismal-forming disk.   

 \par Due to the higher condensation temperature of sulfur-containing volatile molecules, it has been suggested that measuring a C/S and S/O ratio could better constrain both formation location and mechanism \citep{Crossfield_2023}. However, the presence of sulfurous compounds can only be detected in wavelengths $>$3 \micron, opening the door for JWST to make a precise measurement of sulfur abundances of \planet. The measurement of volatiles could distinguish a formation via pebble accretion \citep{Schneider_Bitsch_volatiles_pebbles} versus planetesimal accretion \citep{Pacetti_22_volatiles_planetesimals} at an initial planet formation location of 10 au within the protoplanetary disk, where C/S and O/S are expected to be $\sim1.5$ to 30x enhanced relative to solar under a pebble accretion scenario \citep{Crossfield_2023}. Sulfur is expected to be bound in H$_{2}$S at temperatures of T $\geq$ 1000 K, which is detectable with JWST from 3 - 5 \micron  \citep{2023Natur.617..483T,Fu_2024_h2s}. A detection super-solar C/S and O/S would require an attempt through theoretical modeling of core accretion to explain the formation of ultra-massive planets from 20-30 \Mj at close-in separations. If a solar or sub-solar C/S and O/S ratio are measured, there is tension with either a 20-30 \Mj companion forming via disk fragmentation at 10 au separation, or formation further out in the disk, along with the planet-mass companion, c, and joint migration inward while leaving an intact debris disk extending from $\sim$30 to 180 au. In either scenario, more complex modeling work will need to be done to explain the formation of the HD 206893 system.

\par Understanding the orbital stability of this system is key to determining its formation and evolution. We used the available orbital information reported in \citealt{Hinkley_2023_c} along with novel relative RVs from HD 206893 B to assess the system stability at 1 Myr and 1 Gyr timescales. We find that there are no strong orbital constraints required to make the system stable at 1 Myr (stable $\sim$ 95.8\% of the time). However, for 1 Gyr, there is a strong preference for a larger semi-major axis and near-circular orbits for both companions. Our orbit fit tests for individual companions presented on Appendix \ref{orbit_appendix} confirm that this result remains true regardless of prior choice or mass of HD 206893 B. It is likely as well that the objects' orbital parameters are not fully constrained, which will affect the overall stability assessment, in particular for longer timescales (e.g., \citealt{DoO_orbit_fit_2023}).
\par



\section{Conclusion}\label{sec:conclusion}

\par We have obtained two epochs of high-resolution spectra (R $\sim 35,000$) on the outer ultra massive planet of the HD 206893 system in K-band with Keck/KPIC. We have cross-correlated the spectra with a full atmospheric template and detect the companion at $> 10 \sigma$ significance. This is the first characterization of HD 206893 B at high spectral resolution. Using the extracted spectra within a forward-modeled, Bayesian framework, we infer the most likely bulk atmospheric parameters and radial velocity of the companion. We attempt to make a conclusive projected spin measurement of the companion, but due to the apparently low rotation rate of the companion, we can only establish a $3-\sigma$ upper limit on \vsini of 15.5 km s$^{-1}$. We used a variety of atmospheric models such as \texttt{BT-Settl} \citep{allard_2012_btsettl} and \texttt{PHOENIX} \citep{Barman_2011,Brock_2021} to attempt to cross-validate our bulk-property inferences between model grids, and re-analyzed archival low-resolution Gemini Planet Imager data \citep{Ward-Duong_GPI_spec} using both forward-modeling techniques with the same atmospheric grid parameters and a free-retrieval framework provided by \texttt{petitRADTRANS} to confirm a lower effective temperature, lower log$(g)$ best-fit model and a physically-motivated radius. 

\par We determined an age based on atmospheric evolutionary modeling of 112\plusminus{36}{22} Myr which is consistent with age estimations from previous works at $\sim$150 Myr derived from luminosity and dynamical mass. Using the same atmospheric evolutionary models, we also derived a mass and radius estimate for this companion of $22.7$ \plusminus{2.5}{1.7} M$_{Jup}$ and $1.11 \pm 0.03$ R$_{Jup}$ respectively, which are consistent with VLT/GRAVITY astrometric mass measurements. We believe that analyzing our high-resolution KPIC data in tandem with the stellar evolution models from CBPD23 combined with the latest \texttt{PHOENIX} model grid yield tighter constraints on molecular abundances and bulk atmospheric properties, as well as access to important information like the C/O ratio that can inform the system's formation history.

\par Using our best-fit temperature and \logg model fit, we also determined the best-fit C/O ratio of the companion. This is the first measurement of the C/O ratio of HD 206893 B using a forward modeling framework and yields a solar C/O ratio and metallicity, consistent with either a pebble accretion + runaway gas accretion formation framework or a disk fragmentation formation framework. However, the present separation of the massive companion at 11.62 \plusminus{0.96}{0.88} au is likely not formed via disk fragmentation which is expected to form planets at larger orbital distances of $\sim$100 au \citep{Boley_formation_location_2009}. Migration inward after formation due to mutual torque from the protoplanetary disk is not a likely explanation of the system configuration in this case due to the intact debris disk detected from $\sim$30 to $\sim$180 au \citep{Marino_2020_disk_alma}, as well as the observed solar metallicity, showing no evidence of enhancement from accretion of metals during migration \citep{Helled_2014prpl.conf..643H}. The orbital stability determined from our orbit fitting suggests both planets could have formed at their current locations with no orbital migration necessary. Though formation via core accretion is thought to be most efficient from 5-10 au, population simulations including massive planets formed via core accretion show evidence of these planets tacking out to wider separations where they can be observed via direct imaging \citep{Vigan_Sphere_SHINE_2021}.

\subsection{Future Observations}
\par Given the relatively short orbital period of HD 206893 B, future RV and astrometry measurements will provide a large percentage more orbital coverage of this object and help resolve the degeneracy in its orbit fit. An RV measurement made in 2024 could serve to rule out a higher eccentricity orbit, which has implications for the orbital stability of the system as a whole, as well as population-level statistical measurements of brown dwarf-mass vs. planetary mass companion eccentricity distributions \citep{Bowler_ecc_2020,DoO_orbit_fit_2023}. In addition, longer-wavelength, medium-resolution spectroscopy of \planet \xspace could help to determine the relative ratio of sulfur in the atmosphere. An upcoming JWST Cycle 3 proposal, GO 5485, will use NIRSpec to collect spectra between 3 and 5 $\mu$m with the aim of constraining the C/S ratio from the expected presence of hydrogen sulfide. The C/S ratio promises to be another, more reliable, indicator of formation location than C/O, and will provide further insight into the nature of this extremely massive planetary system.  

\section{Acknowledgments}
\par This research has made use of the NASA Exoplanet Archive, which is operated by the California Institute of Technology, under contract with the National Aeronautics and Space Administration under the Exoplanet Exploration Program.  
\par B.S and Q.K. acknowledge support from NSF grant AST-2046883.  
\par C.D.O. is supported by the National Science Foundation Graduate Research Fellowship Program under Grant No. DGE-2038238.
\par K.H. is supported by the National Science Foundation Graduate Research Fellowship Program under Grant No. 2139433. 
\par J.X. is supported by the NASA Future Investigators in NASA Earth and Space Science and Technology (FINESST) award \#80NSSC23K1434.

Funding for KPIC has been provided by the California Institute of Technology, the Jet Propulsion Laboratory, the Heising-Simons Foundation (grants \#2015-129, \#2017-318, \#2019-1312, \#2023-4598), the Simons Foundation, and the NSF under grant AST-1611623.

\par Some of the data presented herein were obtained at the W. M. Keck Observatory, which is operated as a scientific partnership among the California Institute of Technology, the University of California, and the National Aeronautics and Space Administration. The W. M. Keck Observatory was made possible by the financial support of the W. M. Keck Foundation. The authors wish to acknowledge the significant cultural role that the summit of Maunakea has always had within the indigenous Hawaiian community. The author(s) are extremely fortunate to conduct observations from this mountain. 
\par Portions of this work were conducted at the University of California, San Diego, which was built on the unceded territory of the Kumeyaay Nation, whose people continue to maintain their political sovereignty and cultural traditions as vital members of the San Diego community.

\facilities{Keck:II, Exoplanet Archive}
\software{astropy \citep{astropy:2013,astropy:2018,astropy:2022}, breads (\href{https://github.com/jruffio/breads/tree/main}{https://github.com/jruffio/breads/tree/main}, corner \citep{corner:2016},  efit5 \citep{Meyer_efit5_2012}, emcee \citep{emcee_2013}, h5py (\href{https://www.h5py.org}{https://www.h5py.org}), KPIC Data Reduction Pipeline (\href{https://github.com/kpicteam/kpic_pipeline}{https://github.com/kpicteam /kpic\_pipeline}),  matplotlib \citep{Hunter:2007},  numpy (\href{https://numpy.org}{https://numpy.org}, orbitize! \citep{Blunt_2020_orbitize}, pandas \citep{reback2020pandas}, petitRADTRANS  \citep{Molliere_2019_prt_code}, scipy \citep{2020SciPy-NMeth}, smart  \citep{Hsu_smart_2021ApJS..257...45H}}
\clearpage
\bibliography{KPIC}
\bibliographystyle{aasjournal}

\appendix 
\section{BT-Settl Corner Plot}
\begin{figure*}[ht!]
    \begin{centering}
    \includegraphics[width=0.68\linewidth]{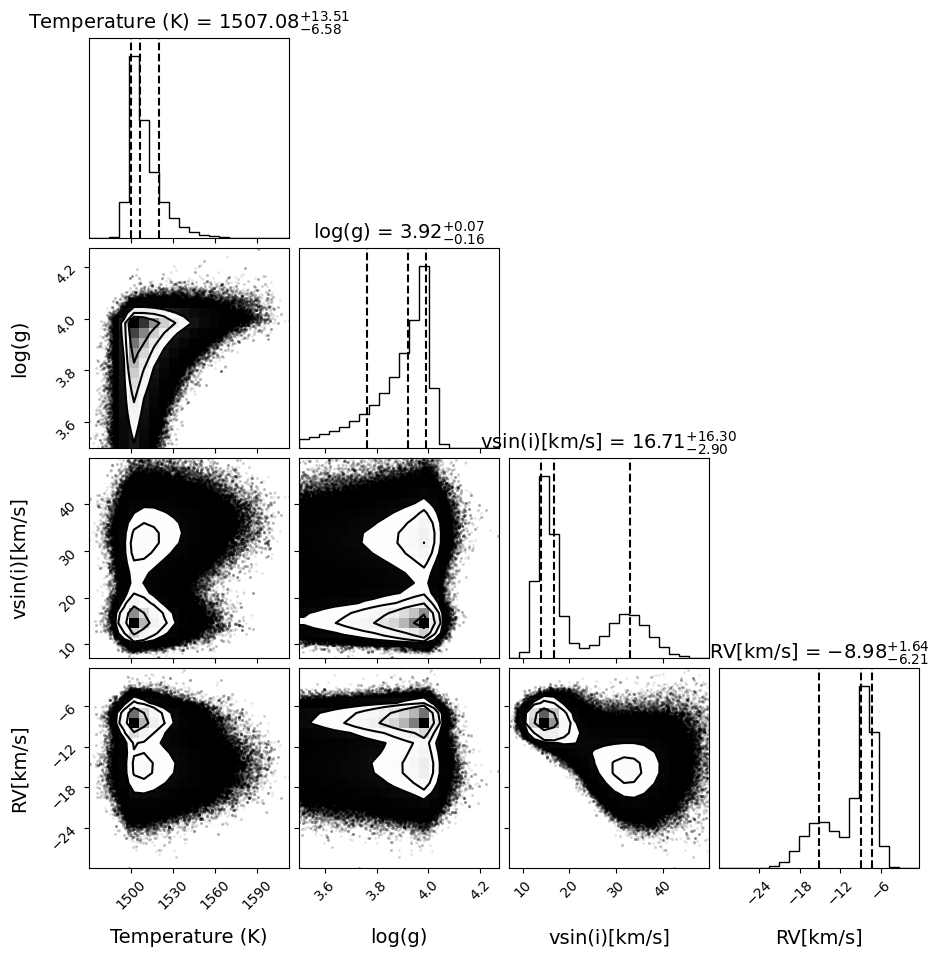}
    \caption{Corner plot for HD 206893 B showing effective temperature, surface gravity, \vsini, and RV using the \texttt{BT-Settl} \citep{allard_2012_btsettl}  model grid and the KPIC \kband spectra. The corner plot shows a bimodal distribution in spin and radial velocity.} 
    \label{fig:BTsettlcornerplot}
    \end{centering}
\end{figure*}

\begin{figure*}[ht!]
    \begin{centering}
    \includegraphics[width=0.8\linewidth]{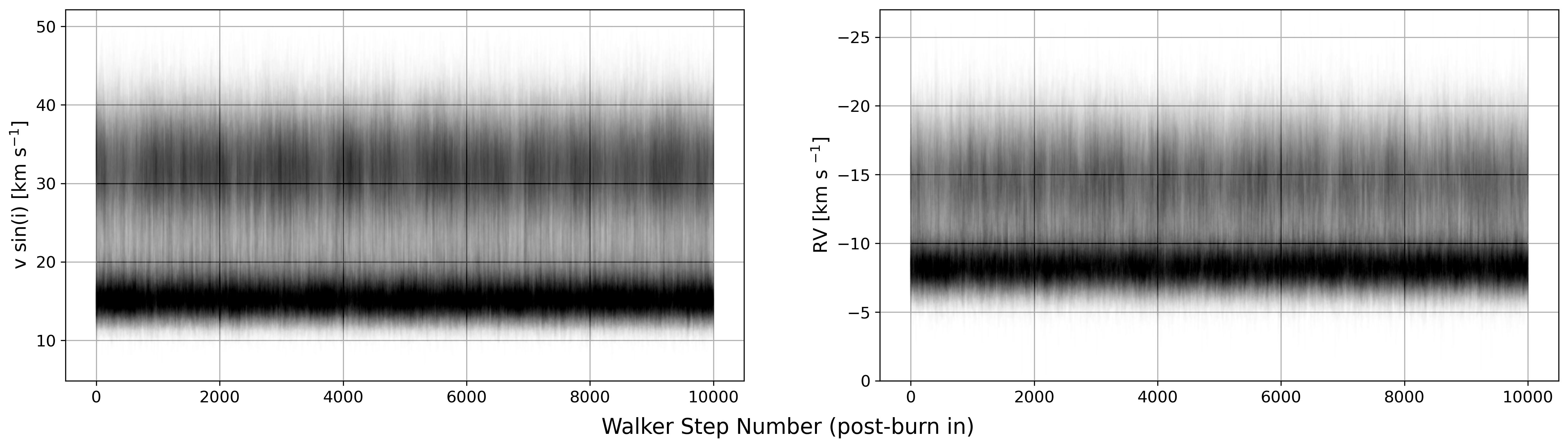}
    \caption{The walker plot for both \vsini and RV from \ref{fig:BTsettlcornerplot}. The MCMC has converged to a bimodal distribution based on the density of the walker traces. }
    \label{fig:btsettlwalkertrace}
    \end{centering}
\end{figure*}
\newpage
\section{\prt Corner Plots}
\begin{figure*}[!hp]
    \includegraphics[width=\linewidth]{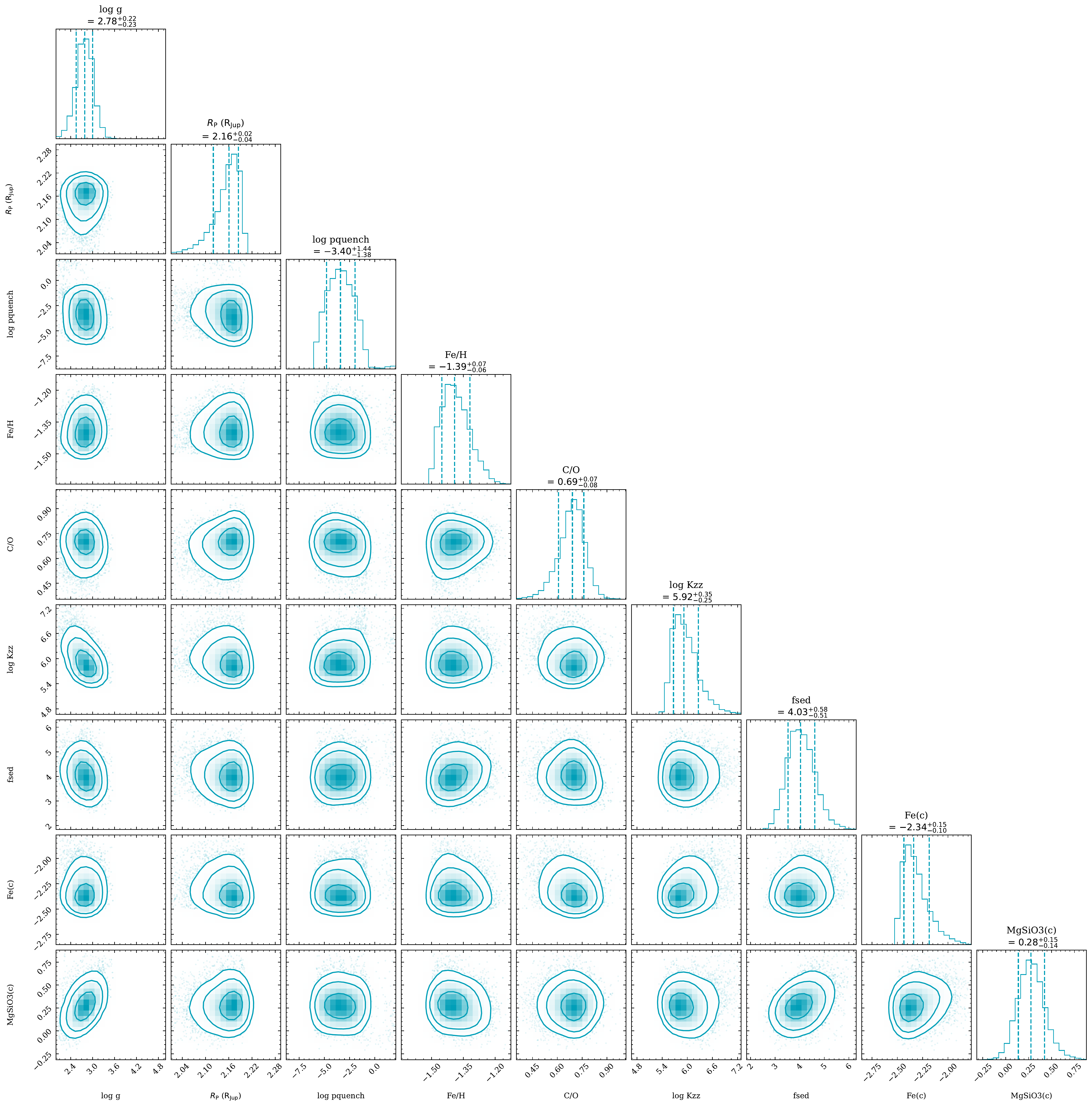}
    \caption{Corner Plot generated by \prt Retrieval fitting GPI \jband, \hband, and \kband data simultaneously from \citealt{Ward-Duong_GPI_spec}.}
    \label{fig:prtcorner_full}
\end{figure*}

\begin{figure*}[!hp]
    \includegraphics[width=\linewidth]{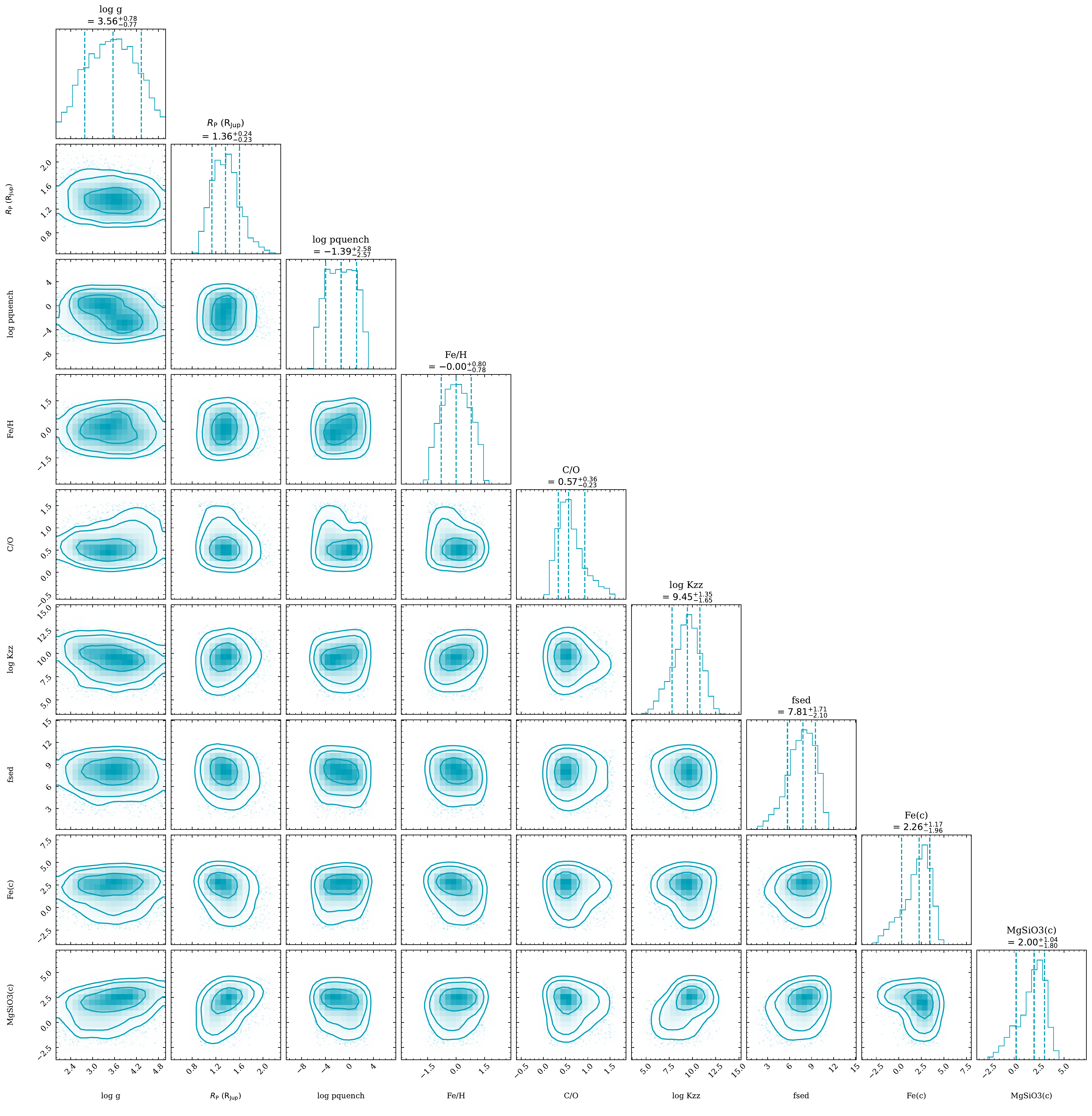}
    \caption{Same as Figure \ref{fig:prtcorner_full}, but only fitting using \jband data.}
\end{figure*}

\begin{figure*}[!hp]
    \includegraphics[width=\linewidth]{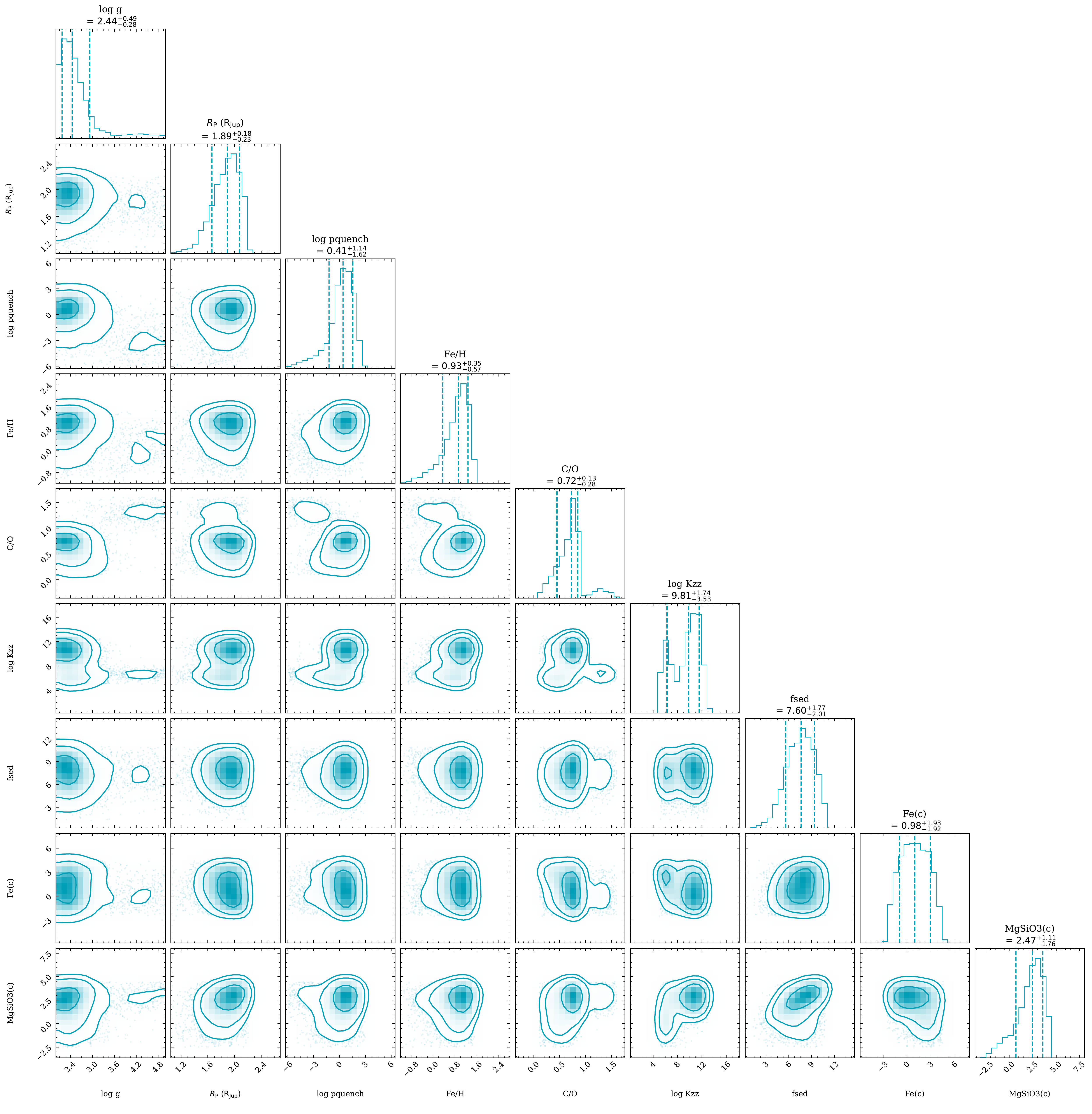}
    \caption{Same as Figure \ref{fig:prtcorner_full}, but only fitting using \hband data.}
\end{figure*}

\begin{figure*}[!hp]
    \includegraphics[width=\linewidth]{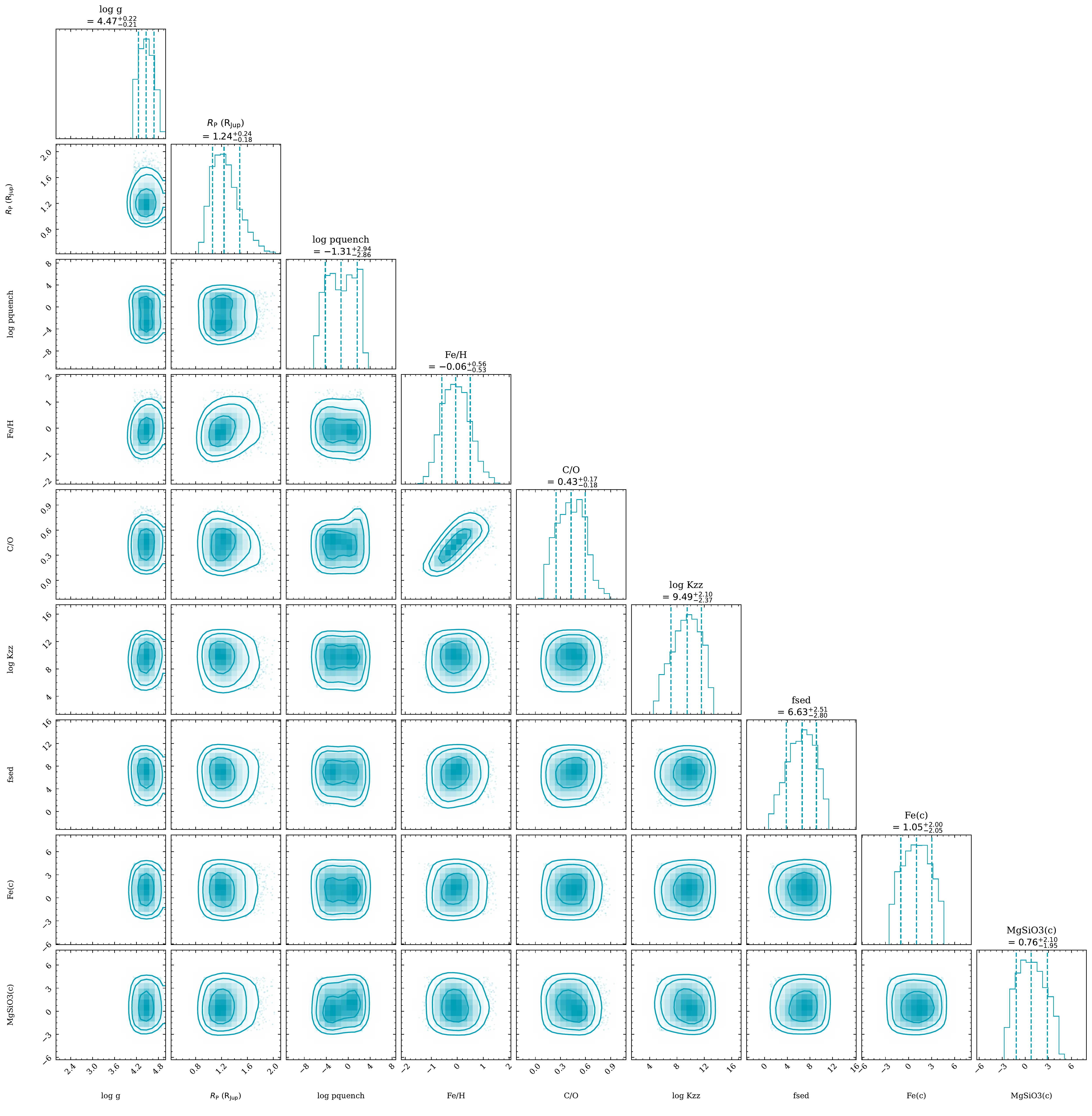}
    \caption{Same as Figure \ref{fig:prtcorner_full}, but only fitting using \kband data.}
\end{figure*}
\newpage
\section{Individual Orbit Analysis} \label{orbit_appendix}

\subsection{HD 206893 B (Alone)}
\begin{figure}
    \centering
    \includegraphics[width = 0.95\linewidth]{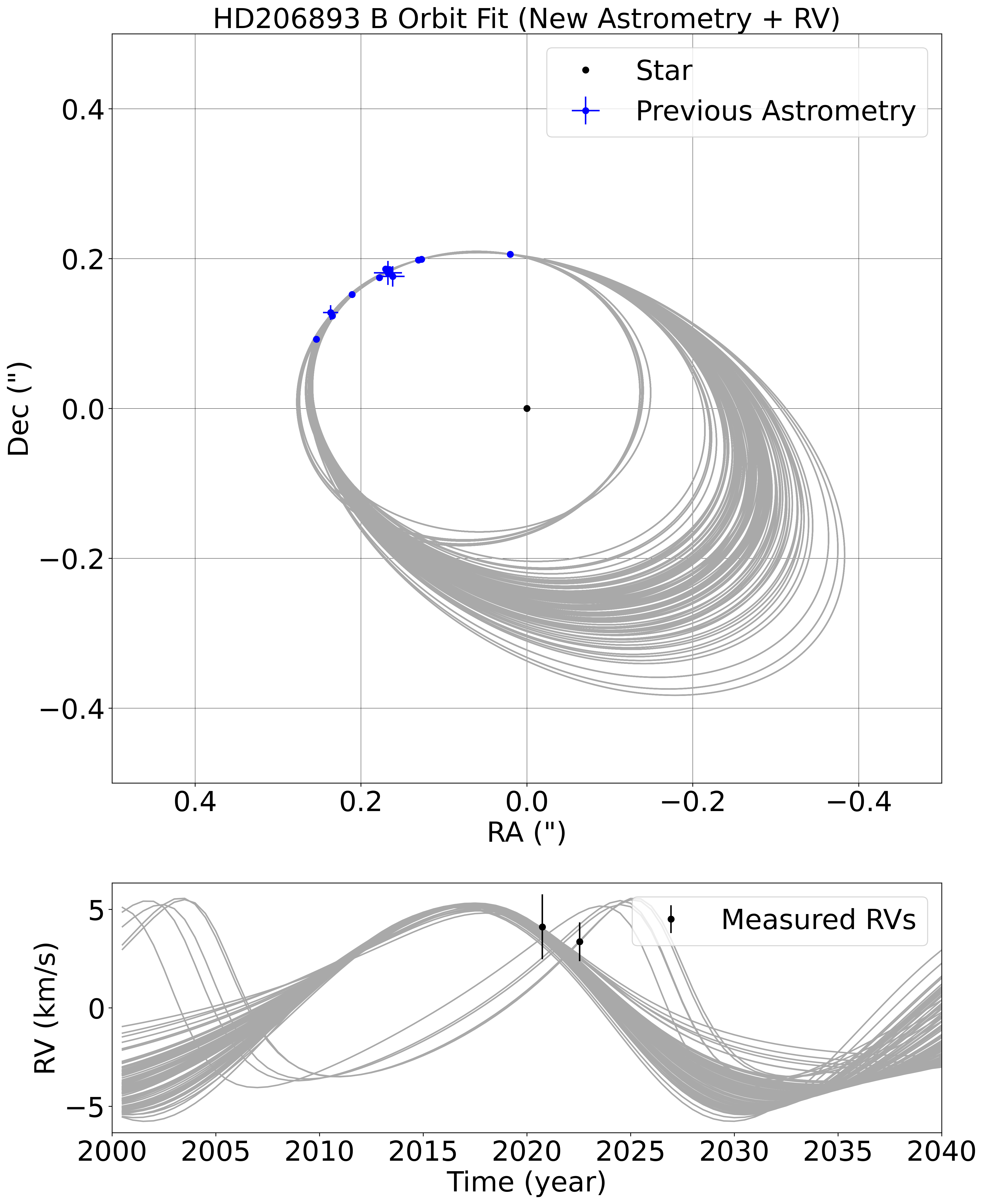}
    \caption{Top: Sampled Orbital fitting of HD 206893 B including astrometry from \citet{Hinkley_2023_c} and Radial Velocity measurements from KPIC in 2020 and 2022 using observable-based priors. Orbits are shown in grey, with the astrometry measurements in blue. Bottom: The Radial velocities of each grey orbit are plotted with KPIC RV measurements.}
    \label{fig:orbit+rv}
\end{figure}

We use the \texttt{Efit5} \citep{Meyer_efit5_2012} software package to fit the published astrometry points \citep{Milli_discovery,Hinkley_2023_c} with the addition of our 2020 and 2022 RV values, but excluding \textit{Gaia} and \textit{Hipparcos} stellar astrometry points. \texttt{Efit5} uses \texttt{MULTINEST} \citep{Feroz_Hobson_multinest_2008,MULTINEST}, nested sampling algorithm based on the principles of Bayesian analysis to fit the data. We used 3,000 live points in the nested sampling algorithm and fit using both the observable-based prior probability distribution as defined in \citealt{O'neil_obs_based_priors_2019} and the one used by \citealt{DoO_orbit_fit_2023}, as well as a uniform prior distribution. The results of the orbital fit are shown in Table \ref{tab:orbits}. The addition of KPIC RV points does not result in better constraints of the eccentricity of the orbit, $e = $ 0.14 \plusminus{0.07}{0.05} when compared to \citealt{Hinkley_2023_c}. The resulting eccentricity is more precise than the joint fit obtained in Section \ref{orbital_analysis}. We find a larger semimajor axis of 11.62 \plusminus{0.96}{0.88} au compared to \citealt{Hinkley_2023_c} and to our joint fits. This orbit fit does not include epicyclic perturbations of HD 206893 c on HD 206893 B. A sub-sample of the 3000 orbits along with the radial velocity values is shown in Figure \ref{fig:orbit+rv}.\par

\subsection{System Stability}
Using the observable-based priors \citep{O'neil_obs_based_priors_2019} implemented in \texttt{Efit5} \citep{Meyer_efit5_2012}, we fit for the orbit of HD 206893 B. We then perform a stability analysis of the system. We treat the orbit of HD 206893 c in three different ways:
\begin{itemize}
    \item Case 1: With the median orbital solutions reported in \citealt{Hinkley_2023_c}. For this configuration, c has a moderate eccentricity of 0.41 and is non-coplanar with B, with an inclination of 150.9\degree as opposed to 139\degree for B. 
    \item  Case 2: With observable-based priors in \texttt{Efit5} \citep{Meyer_efit5_2012} using the relative astrometry of the companion. For this fit, since there are only three astrometric points, \texttt{Efit5} requires that one of the companion’s parameters is constrained (a completely unconstrained fit would require at least four astrometric points). We choose to constrain the inclination of HD 206893 c to 150.9 degrees, to make it consistent with the fit found by \citealt{Hinkley_2023_c}. We also perform another fit where we constrain the inclination of c to 139\degree, which would make the inclination of HD 206893 c consistent with the median inclination of HD 206893 B.
    \item Case 3: With uniform priors in \texttt{orbitize!} \citep{Blunt_2020_orbitize}. For this fit, we also incorporate stellar RVs, as is done in \citealt{Hinkley_2023_c}.	
    \end{itemize}

\par For all cases, we use the orbital posteriors for both HD 206893 B and HD 206893 c combined with the mass fit for HD 206893 B to assess the stability of the system. We divide the mass of B into 10 bins (from 17 to 37 \Mj), the eccentricity of B into 5 bins (from 0 to 0.4). For each bin, we draw 1,000 random orbit combinations from the posteriors and assess whether they remain stable for 1 Myr using the MEGNO parameter \citep{Maffione_Giordano_Cincotta_2011}. We report stability as a percentage of 1,000 configurations that remain stable. 

\par
\begin{figure}
    \centering
    \includegraphics[width=\linewidth,trim={0 0 3.5cm 0}]{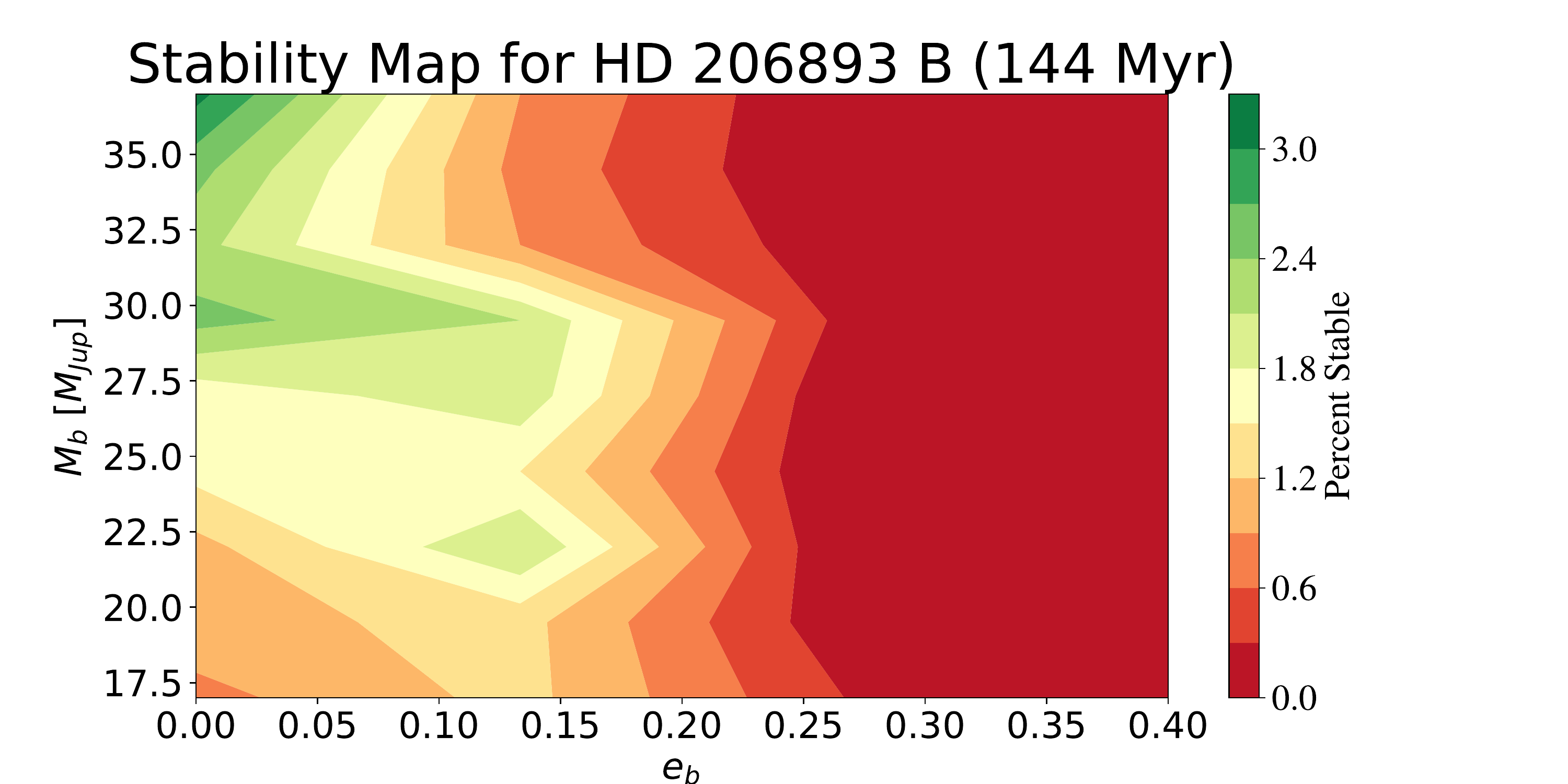}
    \caption{We investigate the stability of \planet \xspace for the retrieved age of the system, 144 Myr. We observe that there are islands of stability, favoring lower eccentricity. Like our mass estimates from evolutionary modeling, we find that 22 \Mj masses and 30 \Mj mass objects appear to have the highest percentages of stable orbits over the life of the system, with the most stable configuration being a 35 \Mj companion on a perfectly circular orbit.}
    \label{fig:fullsystemagestability}
\end{figure}

 For Case 1 specifically, we let the system evolve for the determined age of 144 Myr and find that although the configuration of Case 1 is possible, its maximum stability is 3\% (i.e., only 30/1000 configurations survive in each bin). Stability results are shown in Figure \ref{fig:fullsystemagestability}. It is also apparent from the plot that lower eccentricities of b are preferred, while the higher masses are loosely preferred.

\par
\begin{figure*}
    \centering
    \includegraphics[width=\linewidth]{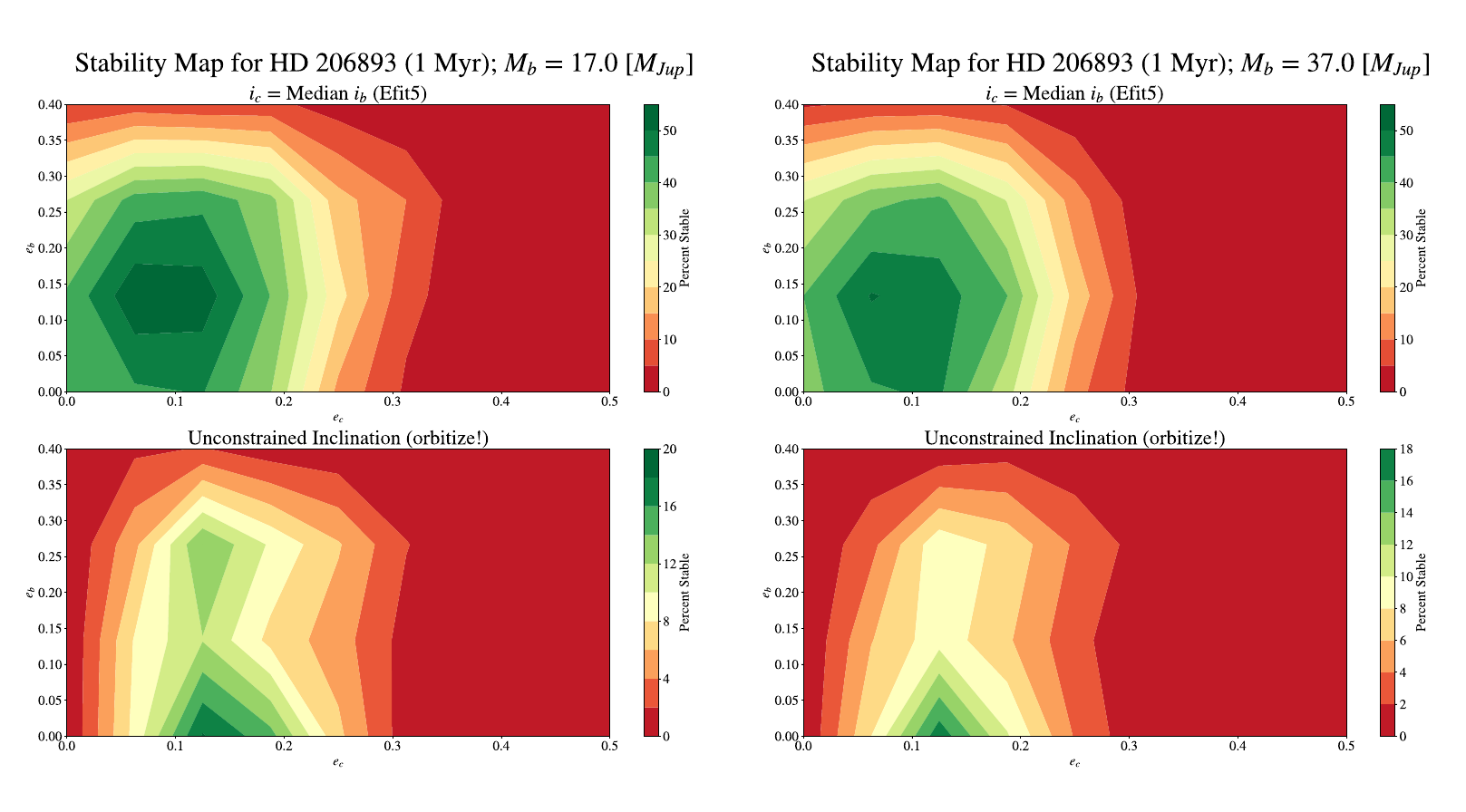}
    \caption{While jointly fitting HD 206893 B and c's orbits, we find an insensitivity to the mass of B. To illustrate this effect, we select the MEGNO plots for Mass$_{B}$ = 17.0\Mj and Mass$_{B}$ = 37.0\Mj using the orbital parameter determinations from both \texttt{Efit5} and \texttt{orbitize!} after letting the planets orbit for 1 Myr. For any chosen value of the mass of B, the most stable systems require an orbital eccentricity of B of $\sim 0.14$ and a low value of eccentricity for c of $<0.2$ with the most stable orbits at $e_c \sim 0.1$.}
    \label{fig:megno_mass_comp}
\end{figure*}

 For Cases 2 and 3, we repeat the initial procedure but also bin the eccentricity posteriors of c into 10 bins (from 0 to 0.5), e.g., $0.0 < e_b < 0.1, 0.25 < e_c < 0.31, 17 \Mj< m_b < 29.5 \Mj$.  This yields 500,000 possible orbit combinations from the objects' posteriors, with 1,000 per bin (and a total of 500 bins). We then assess how many orbits in each bin remain stable for 1 Myr, with this duration chosen for computational considerations. We find in both test cases iterating over the mass of HD 206893 B yields similarly shaped contour plots, showing that the mass is not as significant of a factor in the stability of the system as the eccentricity of the objects. This trend is shown in Figure \ref{fig:megno_mass_comp}, comparing the most extreme cases for the mass of b; 17.0 \Mj in the low mass case and 37 \Mj in the high mass case.  

\par
\begin{figure*}
    \centering
    \includegraphics[width=\linewidth]{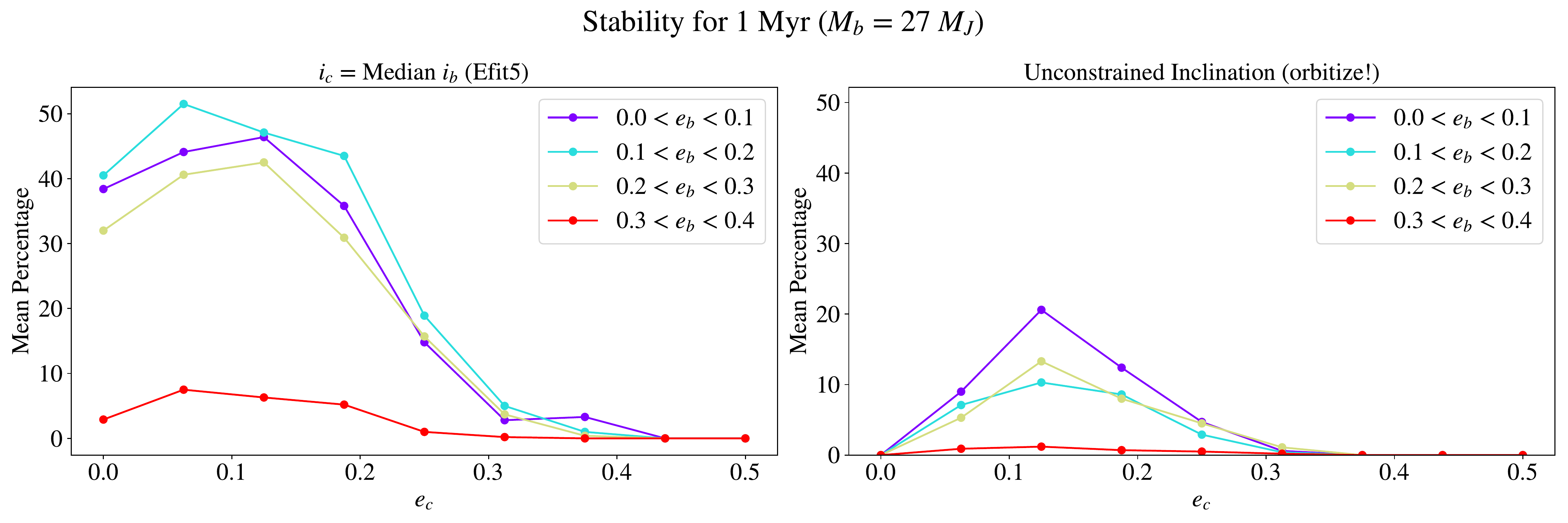}
    \caption{The ``survivability" of coplanar vs non-coplanar systems. When HD 206893 B and c are co-planar, we see a higher maximum percentage of stable systems after 1 Myr with a lower value of $e_c$ and higher value of $e_B$ at $0.1 < e_B < 0.2$.}
    \label{fig:eccentricity+inclination}
\end{figure*}

 We also find that in both cases the system is unlikely to remain stable if the eccentricity of both objects are above 0.4, which is the median value of eccentricity for HD 206893 c found by \citealt{Hinkley_2023_c}. The system is more likely to remain stable when the c has the same inclination as b. Overall, the system's stability appears to favor lower eccentricities for both objects, and to favor co-planarity between the two objects, as demonstrated in Figure \ref{fig:eccentricity+inclination}. This is in agrement with our analysis from the joint orbit fit presented in Section \ref{orbital_analysis}.

\begin{deluxetable*}{cc}
\tablecaption{Orbit Fit Solutions (HD 206893 B)} \label{tab:orbits}
\tablewidth{20pt}
\tablecolumns{2}
\tabletypesize{\scriptsize}
\tablehead{\colhead{Parameter} & \colhead{Observable Prior}}
\startdata
Period (yr) & 34.5 $\pm$ 4.4 \\
Eccentricity & 0.14 \plusminus{0.07}{0.05} \\
$T_0$ (yr) & 2085.1\plusminus{8.5}{30.5} \\
Inclination ($^\circ$) & 139.0\plusminus{2.4}{1.9} \\
$\Omega$ ($^\circ$) & 59.1\plusminus{2.9}{2.2} \\
$\omega$ ($^\circ$) & 50\plusminus{26}{12} \\
\enddata
\tablecomments{The orbital basis of \texttt{Efit5} is different from that of \texttt{orbitize!}.}
\end{deluxetable*}

\end{CJK*}
\end{document}